\documentclass[%
reprint,
showpacs,preprintnumbers,
 amsmath,amssymb,
 aps,
pra,
floatfix,
]{revtex4-1}

\usepackage{graphicx}
\usepackage{subfigure}
\usepackage{dcolumn}
\usepackage{bm}

\def\v{\vskip 12truept}

\def\gl{\gamma_\lambda}
\def\gln{\gamma_{\ell n}}
\def\gn{\gamma_n}

\def\glnp{\gamma_{\ell n^\prime}}

\def\gm{\gamma_m}
\def\lnn{\lambda_n}

\def\ddr{{\partial\over \partial r}}

\def\ddrsq{{\partial^2\over \partial r^2}}
\def\ddtsq{{\partial^2\over \partial t^2}}
\def\vL{\vec L\cdot}
\def\vLg{\vec L\cdot\grad\times}
\def\vrp{\vec r^{\,\prime}}

\def\dom{{\delta\omega}}
\def\grad{\vec \nabla}
\def\lap{\nabla^2}
\def\vr{\vec r}
\def\vrt{\vec r,t}
\def\vE{\vec E}
\def\vB{\vec B}
\def\vD{\vec D}
\def\vEEMlmn{\vec E^{(EM)}_{\ell m n}}
\def\vBEMlmn{\vec B^{(EM)}_{\ell m n}}
\def\vEEMlmpnp{\vec E^{(EM)}_{\ell m^\prime n^\prime}}
\def\vBEMlmpnp{\vec B^{(EM)}_{\ell m^\prime n^\prime}}
\def\vEEMlpmpnp{\vec E^{(EM)}_{\ell^\prime m^\prime n^\prime}}
\def\vBEMlpmpnp{\vec B^{(EM)}_{\ell^\prime m^\prime n^\prime}}
\def\vP{\vec P}
\def\vcE{\vec {\cal E}}
\def\vcP{\vec {\cal P}}
\def\vEp{\vec E^{(+)}}
\def\vBp{\vec B^{(+)}}

\def\vPp{\vec P^{(+)}}

\def\cE{{\cal E}}

\def\cElml{{\cal E}_{\ell m\lambda}}

\def\cElmn{{\cal E}_{\ell mn}}
\def\vElml{{\vec E}_{\ell m\lambda}}
\def\vBlml{{\vec B}_{\ell m\lambda}}

\def\cPlml{{\cal P}_{\ell m\lambda}}

\def\vDglml{{\vec D}^>_{\ell m\lambda}}
\def\vDllml{{\vec D}^<_{\ell m\lambda}}
\def\cP{{\cal P}}
\def\mus{|\vec \mu|^2}

\def\sumlm{\sum_{\ell,m}}
\def\Lm{{\ell m}}
\def\lml{{\ell m\lambda}}
\def\vOm{\vec \Omega}
\def\vOmp{\vec \Omega^\prime}
\def\p{\partial}
\def\ddt{{\partial\over \partial t}}

\def\vrp{\vec r^{\,\prime}}

\def\bdo{\beta_{\delta\omega}}
\def\gdo{\gamma_{\delta\omega}}
\def\vr{\vec r}
\def\vrt{\vec r,t}
\def\vrt{\vec r,t}
\def\vE{\vec E}
\def\vP{\vec P}
\def\vEEMlmn{\vec E^{(EM)}_{\ell m n}}
\def\vEEMlpmpnp{\vec E^{(EM)}_{\ell^\prime m^\prime n^\prime}}
\def\vcE{\vec {\cal E}}
\def\vcP{\vec {\cal P}}
\def\vEp{\vec E^{(+)}}

\def\vPp{\vec P^{(+)}}

\def\cE{{\cal E}}
\def\cP{{\cal P}}
\def\mus{|\vec \mu|^2}

\begin{document}
\preprint{PraGla05}

\title{Coherent Radiation by a Spherical Medium of Resonant Atoms}
\thanks{Submitted to Physical Review A}

\author{Sudhakar Prasad}
\email{sprasad@unm.edu}
\affiliation{Department of Physics and Astronomy\\ University of New Mexico\\ Albuquerque, New Mexico  87131}

\author{Roy J. Glauber}
\affiliation{Lyman Laboratory of Physics\\ 
Harvard University\\ Cambridge, Massachusetts 02138} 
\date{Spetmeber 26, 2010}
\begin{abstract}
Radiation by the atoms of a resonant medium 
is a cooperative process in which the medium participates as 
a whole. In two previous papers \cite{PG00,GP00}, we treated this problem for the case
of a medium having slab geometry, which, under plane wave 
excitation, supports coherent waves that propagate in one dimension. 
We extend the treatment here to
the three-dimensional problem, focusing principally on the 
case of spherical geometry. By regarding the radiation field as
a superposition of electric and magnetic multipole fields of different
orders, we express it in terms of suitably defined scalar fields. The latter fields possess a
sequence of exponentially decaying eigenmodes corresponding to each multipole order.
We consider several examples of spherically symmetric initial excitations of a sphere.
Small uniformly excited spheres, we find, tend to radiate superradiantly, while 
the radiation from a large sphere with an initially excited inner core exhibits temporal oscillations that result from
the participation of a large number of coherently excited amplitudes in different
modes. The frequency spectrum of the emitted
radiation possesses a rich structure, including a frequency gap for large spheres
and sharply defined and closely spaced peaks
caused by the small frequency shifts and even smaller decay rates 
characteristic of the majority of eigenmodes. 
\end{abstract}


\maketitle

\section{\label{I} INTRODUCTION}
A quantum of light emitted within a finite medium made of identical atoms cannot easily
escape. It may suffer a large and undeterminable number of coherent absorption and
reemission processes before reaching the boundary of the medium, and even there 
is subject to the hazard of internal reflection. The emission of a quantum by
a single atom within such a medium, in other words, is inevitably a process in which
the entire medium partakes coherently. When more than one atom is excited
initially, they tend to radiate cooperatively and that sort of
coherent emission process is often referred to as superradiance \cite{Dicke54,AC70,BL76,GH76,GH82,PG85}. The time 
dependence of the energy emitted by the medium in all these cases may differ
considerably from the exponential decay characteristic of the isolated atom,
and its frequency spectrum may differ substantially from the familiar Lorentzian
form. In the present paper we address ourselves to the analysis of these effects 
for the important case in which the medium is spherical in shape and all the
inhomogeneities that destroy coherence are assumed negligible.

The medium we envisage consists of identical atoms, all with the same
electric dipole resonance at a (renormalized) frequency $\omega_0$. Since
they are never more than weakly excited, the atoms may be replaced, in effect,
by harmonic oscillators of the same frequency. We assume them to be distributed
densely enough that many are present in each cubic wavelength $(2\pi c/\omega_0)^3$,
and smoothly enough to permit treating the medium as a continuum. We have called
this idealized model of a resonant and isotropically polarizable medium {\it polarium}, and have
discussed a number of its behaviors in two previous papers, I \cite{PG00} and II \cite{GP00}. 
In those papers, we discussed the emission and propagation of radiation in the 
essentially one-dimensional context of parallel slab geometry. The plane waves
that are radiated, we showed, can be regarded as a superposition of contributions
from a sequence of mutually orthogonal polarization modes that decay exponentially
and have easily calculated properties.

The modal decomposition of the polarization in the one-dimensional medium revealed
in I a complex structure for the time dependence of its decay. An oscillatory
exchange of energy takes place, in effect, between the coherently coupled 
radiation and polarization fields. The radiated spectrum consequently exhibits
an elaborate structure of narrow peaks and dips corresponding to mutually
interefering resonant amplitudes contributed by the various modes. 
A gap is also present in the spectrum, 
corresponding to a band of frequencies in which the implicit dispersion law
of the medium suppresses propagation. Not surprisingly, these features are also found to 
play an important role in the discussion of reflection and transmission of
an externally incident plane wave by a slab shaped medium that we undertook in II. 
By changing the angle of incidence and the polarization direction of the
incident wave, we could tune and alter the spectral dependences of the reflection
and transmission coefficients in predictable ways. 

One of our early tasks in analyzing the problem in spherical geometry will be to find
the appropriate set of exponentially decaying polarization modes. These must
obey conditions that assure the transverse character of the radiated fields. 
In I, transversality was easily secured by dealing only with fields uniform over planes
perpendicular to the axis of propagation. In II, where plane waves could be
obliquely incident upon the slab shaped medium, a more careful treatment
of the separated Cartesian field components was required since induced
surface charges and currents complicate the boundary conditions.

For the spherical geometry we find it particularly convenient to introduce two 
familiar angular momentum projection operators to separate the full
vector field problem into its electric and magnetic multipole parts. 
Each of these parts can be further resolved into a succession of spherical
harmonic components. The magnetic multipole fields separated in this way
automatically obey the required transversality conditions. The electric
multipole fields, on the other hand, still require some consideration of
surface charges and currents in order to secure transversality. The most
convenient feature of the angular momentum decomposition is that
it involves only scalar functions which are in effect the radial components of the electric
and magnetic fields. We may express these components in terms of mode 
functions that decay exponentially with time in each spherical harmonic order.

We begin in Sec. II with the elementary example of radiation from
a small uniformly excited sphere. This problem, which can be solved directly
without the use of angular momentum operators, provides physical
insights into the coherent radiation problem, which are useful
in treating the problem of larger spheres. In Sec. III, we present
the equations of motion for the radiation field and polarization. The 
complications of maintaining a transverse displacement field everywhere
are handled here by means of the two angular momentum projection operators. Their use
leads to equations of motion for the electric and magnetic 
fields and for their polarization sources. The magnetic multipole radiation,
which is analytically simpler to treat, is discussed in Sec. IV, while a full treatment of the 
electric multipole radiation is begun in Sec. V. The 
magnetic multipole radiation field obeys simple outgoing-wave boundary conditions in 
each angular momentum order. We derive the eigenvalue equation that results 
from these boundary conditions and briefly discuss certain important
properties of the exponentially decaying eigenmodes. The electric multipole radiation 
problem requires somewhat more involved outgoing-wave boundary conditions that 
can be most simply treated by considering the differential equations obeyed
by the electric and displacement fields inside and outside the medium.
When the medium excitation is restricted to 
initial polarizations that are oriented along a fixed direction and have
a radially symmetric but otherwise arbitrary amplitude, the spherical medium,
regardless of its radius, radiates as a pure electric dipole, as we show in Sec.~V.
In the same section,
the associated electric-dipole eigenvalue problem is developed in the more general
context of electric multipoles of arbitrary order, and the eigenvalue equation
is analytically solved for the eigenvalues in several important limiting cases. 
In Sec.~VI, we establish the orthogonality of electric multipole modes of an
arbitrary order. In Sec.~VII we discuss the temporal and spectral 
characteristics of the electric dipole radiation from an interesting example of a spherical
system, one in which a uniformly 
excited spherical core is surrounded by an initially unexcited spherical shell.
Finally, in Sec. VIII, some concluding remarks about the problem of coherent
transport of resonant excitations treated here are presented.
 
\section{\label{II} AN ELEMENTARY EXAMPLE: RADIATION FROM A SMALL UNIFORMLY EXCITED 
SPHERE}

It will be useful to begin our analysis by discussing the radiation by a 
spherical medium of radius $R$ much smaller than the reduced wavelength $1/k_0
=c/\omega_0$, {\it i.e.,} $\beta \equiv k_0R <<1$. The problem is simple 
enough to afford elementary access.
It will furnish a valuable example for later reference.

The atoms of the medium we call {\it polarium} are assumed to be distributed with a uniform density
$n_0$. The transition matrix elements of their electric dipole moment vectors
$\vec \mu$ are assumed to be randomly oriented so that the medium is isotropic,
and its induced electric polarization is always parallel to the inducing field. Then, as we
have shown in deriving Eq.~(9) of I, the positive frequency part of the 
polarization field $\vPp(\vrt)$ for such a medium
is driven by the positive frequency part of the electric field $\vEp(\vrt)$ through the 
relation
\begin{equation}
\label{e1}
\left(\ddt +i\omega_0\right)\vPp(\vrt)={in_0\mus\over 3\hbar}\vEp(\vrt).
\end{equation}
In the absence of the electric field, the polarization
varies in time at any point as $\exp(-i\omega_0t)$.

By expressing $\vEp$ and $\vPp$ in terms of their 
slowly-varying envelopes $\vcE$ and $\vcP$:
\begin{equation}
\label{e2}
\vEp(\vrt)=\vcE(\vrt)e^{-i\omega_0 t},\ \ 
\vPp(\vrt)=\vcP(\vrt)e^{-i\omega_0 t},
\end{equation}
we may rewrite Eq.~(\ref{e1}) as 
\begin{equation}
\label{e3}
\ddt \cP(\vrt)={in_0\mus\over 3\hbar}\cE(\vrt).
\end{equation}

We shall assume that the polarization is spatially uniform and given by $\hat z P(t)
\exp(-i\omega_0 t)$, where $\hat z$ is a unit vector along the $z-$axis.
In that case, the fields generated by the uniform polarization, although
rapidly oscillating in time, have the familiar electrostatic spatial 
dependence within the near field zone. Thus the electric field within
the sphere is uniform and parallel to the polarization. 
The magnetic field within the medium is of relative order $k_0R<<1$,
and is thus negligible in this, the long wavelength limit.

Such a uniformly polarized small sphere decays superradiantly, as we shall see
presently. If $\lambda_0$ is the exponential decay constant for this mode,
then the electric field within the medium, which we denote by $\vE^<$, and the polarization 
$P_0\exp(-\lambda_0 t)$ are formally related, according to Eq.~(\ref{e3}), by
\begin{equation}
\label{e4}
\vE^<=\hat z{3i\hbar\lambda \over n_0\mus}P_0 e^{-\lambda_0 t}
e^{-i\omega_0 t}.
\end{equation}
Outside the medium, the electric field $\vE^>$ is that of a point dipole 
$\vec p_0$, equal in value to the dipole moment of the sphere, 
\begin{equation}
\label{e4p}
\vec p_0=\hat z P_0 {4\pi\over 3} R^3, 
\end{equation}
and located at its center.
In the near field zone, the electric field thus assumes the familiar elecrostatic
spatial dependence
\begin{equation}
\vE^>={3(\vec p_0\cdot \hat r)\hat r-\vec p_0\over 4\pi r^3}e^{-\lambda_0 t}
e^{-i\omega_0 t}.
\end{equation}

The eigenvalue $\lambda_0$ is determined by requiring that in the long-wavelength limit the electric
field in the interior of a uniformly polarized small sphere 
be minus a third of its polarization, $\vec E^< = -(1/3)\vec P$. 
In view of Eq.~(\ref{e4}), this requirement yields the following expression for $\lambda_0$:
\begin{equation}
\label{e5}
\lambda_0=i{n_0\mus \over 9 \hbar}.
\end{equation}

Because the eigenvalue (\ref{e5}) is purely imaginary, it
represents merely a frequency shift for the field and polarization. The 
sphere does not radiate in the limit $\beta\to 0$.
For small but finite $\beta$, the sphere does, in fact, radiate.
To calculate the decay rate ${\rm Re} \lambda_0$, we resort
to an analysis based on the rate at which a uniformly polarized sphere,
with its polarization oscillating at frequency $\omega_0$, radiates
energy. A small sphere of this kind radiates
as an oscillating point dipole \cite{Jackson1} at the time-averaged rate 
\cite{footnote1}
\begin{equation}
\label{e6}
W={c |\vec p_0|^2 k_0^4\over 3\pi}.
\end{equation}

Equivalently, the rate at which energy is lost from the dipoles of the 
spherical medium must be equal to the rate at which work is done by them on 
the field. The rate of work done by the $i$th dipole of dipole moment
$\vec p_i$, when averaged over 
the fundamental period of oscillation, is just $-2{\rm Re}\,
\dot{\vec p_i^*} \cdot \vec E$, where $\dot{\vec p_i}$ is the time
derivative of the dipole moment. Therefore the rate at which work is done by the sphere is 
$-2 {\rm Re}\,
\dot{\vec P^*} \cdot \vec E (4\pi R^3/3)$. Since the polarization and
field both oscillate at a frequency close to $\omega_0$, the preceding
expression is essentially the same as
\begin{equation}
\label{e7}
W={8\pi R^3\over 3}\omega_0{\rm Im}\,\vec P^* \cdot \vec E.
\end{equation}
When the relation (\ref{e1}) between the field and polarization is used to 
eliminate $\vE$ from Eq.~(\ref{e7}), the 
resulting expression for $W$ is
\begin{equation}
\label{e8}
W={8\pi R^3\omega_0}{{\rm Re}\,\lambda_0\over (n_0\mus/\hbar)} |\vec P|^2.
\end{equation}
An explicit expression for the decay rate ${\rm Re}\,\lambda_0$ is now
obtained by equating Eqs.~(\ref{e6})) and (\ref{e8}) and using Eq.~(\ref{e4p}), 
\begin{equation}
\label{e9}
{\rm Re}\,\lambda_0= {2\over 27} {n_0\mus\over \hbar}\beta^3.
\end{equation}
The rate (\ref{e9}) at which a small uniformly polarized sphere decays radiatively 
may be expressed as the product of the number of atoms, $N=n_0 (4\pi R^3/3)$, and 
the Wigner-Weisskopf intrinsic decay rate of each atom, $\tau^{-1}=\mus
k_0^3/(18\pi\hbar)$:
\begin{equation}
\label{e10}
{\rm Re}\,\lambda_0={N}\tau^{-1}.
\end{equation}

Each dipole in an assembly of $N$ identically
prepared and coherently coupled atomic dipoles emits radiation at a rate
that is $N$ times the rate with which it would spontaneously radiate when 
isolated from the others. Such enhanced decay rates are characteristic of
superradiant emission, a process that is often considered in the
context of much stronger excitation of the radiating medium, e.g., 
when all of the atoms are fully excited in the initial state. 
In these more general examples the emission process can only be described adequately
by means of nonlinear equations.
The present problem, by contrast, is considerably simpler due to its linearity 
in the fields and polarization.
Because of the coherent initial preparation of the atomic 
dipoles, the essential coherence always remains present in the emission process.

It is worth recalling here that Hartmann and collaborators \cite{FHM72,FH74} have made
an important criticism of Dicke's elementary theory of superradiance in many-atom systems.
They have pointed out that the electric dipole moments induced in different atoms
will interact strongly via the familiar dipole-dipole interactions and lead to
spatially dependent shifts of atomic energy levels. These differing level shifts
can bring about relative dephasing of different parts of the oscillating polarization
distribution and thus some breakdown of the cooperative character of superradiant
emission. We see no evidence of this suppression in the radiative rate given 
by Eq.~(\ref{e10}). Indeed, the way in which we have treated the interaction of 
each atom with the field implicitly includes the effects of all dipole-dipole
interactions. Their total effect does not inhibit superradiance, at least
for the linear problem of radiation from a small uniformly polarized sphere.

\section{\label{III}FORMULATION OF THE GENERAL PROBLEM}

\v 

We now turn to the general problem of radiation by an arbitrary excitation of 
a spherical medium of arbitrary radius.
The resonant interaction of the polarium medium with the 
electromagnetic field is described by Eq.~(\ref{e1}) and the Maxwell wave equation
\begin{equation}
\label{e11}
-\grad\times(\grad\times \vEp)-{1\over c^2} \ddtsq\vEp=
{1\over c^2} \ddtsq\vPp.
\end{equation}
Because of the identity 
\begin{equation}
\label{e12}
-\grad\times(\grad\times \vEp)=\lap \vEp-\grad(\grad\cdot \vEp),
\end{equation}
Eq.~({\ref{e12}) contains an explicit gradient term, $\grad(\grad\cdot \vEp)$,
which enforces the transversality of the total displacement field, $\vec D
=\vE+\vP$. We shall see later that this term greatly
influences the character of the radiation and most particularly when the spheres
are small compared to the wavelength of radiation.

The spherical geometry of the radiation problem is best approached 
by decomposing the radiation field into its electric multipole (EM) and magnetic multipole (MM)
components, and introducing appropriate scalar functions to 
describe them. A simple way to exhibit this decomposition without introducing
the full panoply of vector spherical harmonics is
to use two operators \cite{Jackson1,PG85} that are simply related to the quantum-mechanical
angular-momentum
operator, $\vec L=-i\vr\times\grad$. Let us consider the action of the
operators $\vec L\cdot$ and $\vec L\cdot \grad\times$ on Eqs.~(\ref{e1})
and (\ref{e11}). 
Because both these operators annihilate the gradient term inside the
double curl when identity (\ref{e12}) is used and because they commute with
the Laplacian,
Eq.~(\ref{e11}) simplifies to an inhomogeneous scalar wave equation of the general form
\begin{equation}
\label{e13}
\left(\lap-{1\over c^2} \ddtsq\right)\eta(\vrt)=
{1\over c^2} \ddtsq\phi(\vrt),
\end{equation}
where the symbols, $\phi$ and $\eta$, denote the functions 
\begin{equation}
\label{e14}
\phi = \vec L\cdot \vPp \ {\rm and} \ \eta = \vec L\cdot \vEp) 
\end{equation}
or alternatively
\begin{equation}
\label{e15}
\phi=\vec L\cdot\grad\times \vPp \ {\rm and}\  \eta = \vec L\cdot\grad\times \vEp). 
\end{equation}

That $\vL\vEp$ and $\vLg\vEp$ describe the radial components of the 
magnetic and electric fields and thus the MM and EM
fields, respectively, is immediately evident when the 
vector triple products are rearranged, and the Faraday and Maxwell-Amp\`ere 
laws are introduced as follows: 
\begin{align}
\label{e16}
\vL\vEp\sim &(\vr\times\grad)\cdot \vEp=\vr\cdot (\grad\times\vEp)\nonumber\\
&\sim \vr\cdot \ddt \vBp\sim\vr\cdot \vBp\nonumber\\
\vLg\vEp\sim& \vL \vBp\sim\vr\cdot (\grad\times\vBp)\nonumber\\
&\sim \vr\cdot \ddt \vEp\sim\vr\cdot \vEp.
\end{align}
The final step in each of the two relations in Eq. (\ref{e16}) has employed the
assumption of quasi-monochromaticity, which permits replacing time differentiations
of the positive-frequency parts of the electromagnetic field by the multiplicative factor
$-i\omega_0$. The validity of this assumption is assured by
the resonant character of the radiative interactions in the medium. 

By contrast with Eq.~(\ref{e11}), Eq.~(\ref{e1}) is formally unchanged when the 
spatial operator $\vec L\cdot$ is applied, 
\begin{equation}
\label{e17}
\left(\ddt +i\omega_0\right)\phi(\vrt)={in_0\mus\over 3\hbar}\eta(\vrt),
\end{equation}
where $\phi$ and $\eta$ are given by Eq.~(\ref{e14}).
Under the $\vec L\cdot \grad\times$ operation, however, the 
sharp drop-off at the surface of the otherwise uniform medium density contributes 
to the right-hand side of Eq.~(\ref{e17})
a surface singularity \cite{Jackson2}, which is generated by oscillating 
surface charges. Such surface polarization charges and currents must be present for 
EM radiation for which both the
electric field and polarization have nonvanishing radial components. 
A self-consistent approach that treats these surface
singularities correctly but implicitly is based on matching on the surface the appropriate components
of the electromagnetic field inside the medium to those outside. 

The $\vec L\cdot$ operation
commutes with the density which has only a radial step-function form. 
Eq.~(\ref{e17}) is thus valid both in the interior and on the surface of the medium.
That is to say, for MM radiation, the electric field and polarization
are both purely transverse, and no surface charges are present.

For the problem of radiation from an initially excited medium with no
externally incident fields, Eq.~(\ref{e16}) admits
the familiar sort of retarded integral solution for $\eta$:
\begin{eqnarray}
\label{e18}
\eta(\vrt)=&-&{1\over c^2}\int {\partial^2\over\partial t^{\prime 2}}\phi(
\vr^{\,\prime},t^\prime){\delta\left(t-t^\prime-{|\vr-\vrp|\over c}\right)\over 
4\pi|\vr-\vrp|}d\vrp dt^\prime\nonumber\\
=& -&{1\over c^2}\int {{\partial^2\over\partial t^2}\phi\left(
\vrp,t-{|\vr-\vrp|\over c}\right)\over 
4\pi|\vr-\vrp|}d\vrp.
\end{eqnarray}
By expressing $\eta$ and $\phi$ in terms of their 
slowly-varying envelopes $\cE$ and $\cP$:
\begin{equation}
\label{e19}
\eta(\vrt)=\cE(\vrt)e^{-i\omega_0 t},\ \ 
\phi(\vrt)=\cP(\vrt)e^{-i\omega_0 t},
\end{equation}
and dropping the time derivatives of $\cP$, which may be assumed small,
we may reduce Eq.~(\ref{e18}) to the form
\begin{equation}
\label{e20}
\cE(\vrt)=k_0^2\int \cP\left(
\vrp,t-{|\vr-\vrp|\over c}\right){e^{ik_0|\vr-\vrp|}\over 
4\pi|\vr-\vrp|}d\vrp.
\end{equation}

Because Eq.~(\ref{e17}) fails to include radiating surface currents, 
the description provided by Eqs.~(\ref{e17})) and (\ref{e20}) is not complete for 
EM radiation. But these equations do, however, describe properly MM radiation which has 
no surface sources for a spherical radiator. 

\section{\label{IV}MAGNETIC MULTIPOLE RADIATION AND THE ASSOCIATED EIGENVALUE PROBLEM}

The use of the envelopes (\ref{e19}) in Eq.~(\ref{e17}) leads to the equation of motion for
the polarization multipoles,
\begin{equation}
\label{e21}
\ddt \cP(\vrt)=i{|\vec \mu|^2\over 3\hbar}n_0\cE(\vrt).
\end{equation}
By eliminating the field multipoles $\cE$ between Eqs.~(\ref{e20}) and (\ref{e21}), we
secure the integral equation for the polarization multipole fields
\begin{align}
\label{e22}
\ddt\cP(\vrt)=i{\mus n_0k_0^2\over 3\hbar}\int&{e^{ik_0|\vr-\vrp|}\over 
4\pi|\vr-\vrp|}\nonumber\\
&\times\cP\left(\vrp,t-{|\vr-\vrp|\over c}\right)d\vrp.
\end{align}

The integrand on the right side of Eq.~(\ref{e22}) requires evaluating the envelope
function
$\cP$ at the retarded time $t-|\vr-\vrp|/c$, but if its temporal variation is 
sufficiently slow it remains accurate to neglect the 
retardation in $\cP$ and write
\begin{equation}
\label{e23}
\ddt\cP(\vrt)=i{\mus n_0k_0^2\over 3\hbar}\int{e^{ik_0|\vr-\vrp|}\over 
4\pi|\vr-\vrp|}\cP\left(\vrp,t\right)d\vrp.
\end{equation}
The more important effects of retardation are still retained in the exponential
function in the integrand. This approach, we have called the rapid-transit approximation in I,
assumes only that the slowly varying amplitude $\cP$ does not change appreciably during
the passage time of a wave through the medium.
The approximation is not essential to our treatment of the problem, but it greatly 
simplifies the analysis, and we shall therefore employ 
it in exploring the behavior of the system.

Let us consider an expansion of $\cP(\vrt)$ in spherical harmonics,
defined according to the convention employed in Ref.~\cite{Jackson2}, 
\begin{equation}
\label{e24}
\cP(\vrt)=\sumlm P_\Lm(r,t)Y_\Lm(\vOm).
\end{equation}
Substitution of this form into Eq.~(\ref{e23}), followed by a
use of the identity
\begin{equation}
\label{e25}
{e^{ik_0|\vr-\vrp|}\over 4\pi |\vr-\vrp|}=ik_0\sumlm j_\ell(k_0r^<)
h_\ell^{(1)}(k_0r^>) Y_\Lm(\vOm)Y^*_\Lm(\vOmp),
\end{equation}
and integration over the solid angles $\vOmp$ of the vector $\vrp$,
together with a use of the orthonormality and linear independence of the various 
spherical harmonics yields the result
\begin{align}
\label{e26}
\ddt \cP_\Lm(r,t)=-&{n_0\mus k_0^3\over 3\hbar}\int_0^R \cP_\Lm(r^\prime,t)\nonumber\\
&\times j_\ell(k_0r^<)h_\ell^{(1)}(k_0r^>) r^{\prime 2} dr^\prime.
\end{align}
Here $j_\ell$ and $h_\ell^{(1)}$ are spherical Bessel and Hankel
functions of the first kind, and $r^< (r^>)$ is the smaller (larger) of 
the two radial distances $r,r^\prime$. That each multipole order 
separates from all others is a consequence of the spherical geometry.
All multipoles that are initially unexcited thus remain unexcited at all later times.

We now look for solutions of Eq.~(\ref{e26}) that have a purely exponential time dependence,
\begin{equation}
\label{e27}
\cP_\Lm(r,t)=\cP_\lml(r)e^{-\lambda t}.
\end{equation}
For such solutions, Eq.~(\ref{e26}) reduces to the homogeneous Fredholm integral 
equation:
\begin{align}
\label{e28}
\lambda \cP_\lml(r)={n_0\mus k_0^3\over 3\hbar}\int_0^R& \cP_\lml(r^\prime)\nonumber\\
&\times j_\ell(k_0r^<)h_\ell^{(1)}(k_0r^>) r^{\prime 2} dr^\prime.
\end{align}
This equation contains the radiative boundary condition at the spherical
surface $r=R$
\begin{equation}
\label{e29}
\cP_\lml(r)\ {\buildrel r\to R\over \longrightarrow} \ h_\ell^{(1)}(k_0 r)
\end{equation}
together with the condition that $\cP_\lml$ remains finite at $r=0$.
These conditions restrict the form of the polarization function $\cP_\lml(r)$
and the values of the decay constant $\lambda$. 

The possible values of $\lambda$, the
eigenvalues, form a discrete, infinite set of complex numbers. 
By invoking the symmetry of the kernel of the integral equation (\ref{e28}) under the 
interchange $r\leftrightarrow r^\prime$, we can establish, as in I, both 
the positivity of the real part of each eigenvalue and the orthogonality of
the eigenfunctions,
\begin{equation}
\label{e30}
\int_0^R \cP_\lml(r)\cP_{\ell m\lambda^\prime}(r) r^2 dr \sim 
\delta_{\lambda\lambda^\prime}.
\end{equation}
Equation (\ref{e30}) can be used to define a normalization integral and to
secure an orthonormal set of eigenfunctions,
Further properties of the eigenvalues and eigenfunctions follow from the bilinear
expansion of the kernel in terms of the orthonormal eignfunctions,
as we showed in the context of the one-dimensional problem \cite{PG00}.

To solve explicitly for the eigenvalues $\lambda$ and eigenfunctions $\cP_\lml(r)$, we 
employ the fact that the kernel $j_\ell(k_0r^<)h_\ell^{(1)}(k_0r^>)$
is a Green's function,
\begin{align}
\label{e31}
\left\{ {d^2\over dr^2}+{2\over r} {d\over dr}+\left[k_0^2-{\ell(\ell+1)
\over r^2}\right]\right\}
&j_\ell(k_0r^<)h_\ell^{(1)}(k_0r^>)\nonumber\\
&={i\over k_0r^2}\delta(r-r^\prime).
\end{align}
This enables us to convert Eq.~(\ref{e28}) into a differential form,
\begin{equation}
\label{e32}
\left\{ {d^2\over dr^2}+{2\over r} {d\over dr}+\left[\gl^2
-{\ell(\ell+1)\over r^2}\right]
\right\}\cP_\lml(r)=0,
\end{equation}
where 
\begin{equation}
\label{e33}
\gl=k_0\sqrt{ 1-{in_0\mus/(3\hbar)\over \lambda}}
\end{equation}
may be regarded as the propagation constant for the mode inside the medium.

The most general solution of this equation that remains finite at $r=0$ has
the form
\begin{equation}
\label{e34}
\cP_\lml(r)=Aj_\ell(\gl r).
\end{equation}
The radiative boundary condition (\ref{e29}) at $r=R$ is equivalent to the 
following equality involving logarithmic derivatives:
\begin{equation}
\label{e35}
{{d\over dR}\cP_\lml(R)\over \cP_\lml(R)}= 
{{d\over dR}h^{(1)}_\ell(k_0R)\over h^{(1)}_\ell(k_0R)},
\end{equation}
which immediately leads to the eigenvalue equation
\begin{equation}
\label{e36}
{\gl j_\ell^\prime(\gl R)\over j_\ell(\gl R)}=
{k_0h_\ell^{(1)^\prime}(k_0 R)\over 
h_\ell^{(1)}(k_0 R)},
\end{equation}
where the prime superscript denotes first derivatives
of the respective functions with respect to their arguments.

\section{\label{V}EXCITATIONS OF SPHERICALLY SYMMETRIC AMPLITUDE AND ELECTRIC DIPOLE
RADIATION}

If the initial polarization present in the medium has complete spherical 
symmetry, it must point everywhere in the radial direction with an amplitude
that has no angular dependence. Such polarizations, being purely longitudinal,
cannot radiate at all. The radiation of transverse waves requires that 
the spherical symmetry be broken.

We consider radiation by initial polarizations of the spherical medium 
that have a uniform direction throughout the sphere but spherically
symmetric amplitudes. If we take this direction to be the $z$ axis 
identified by the unit vector $\hat z$, we may write
\begin{equation}
\label{e37}
\vPp(\vr,t=0)=\hat z p(r).
\end{equation}
Since $\hat z$ commutes with $\vec L$ which annihilates any spherically
symmetric function, the scalar product $\vL\vPp$ vanishes initially 
and, according to Eq.~(\ref{e23}),
at all subsequent times as well. Because of Eq.~(\ref{e21}), $\vL\vEp$ therefore
also vanishes at all times. The initial excitation (\ref{e37}) thus cannot
radiate MM fields. 

The operation of $\vLg$ on Eq.~(\ref{e37}), on the other hand, produces a nontrivial
result. Because of the density discontinuity at the boundary, this operation
involves a surface singularity, as we noted earlier. It leads, furthermore, to a 
finite result within the medium, that may be shown after simple algebra
to involve the derivative of the amplitude
$p(r)$,
\begin{equation}
\label{e38}
\vLg\vPp(\vr,0)=2i\sqrt{4\pi\over 3}p^\prime (r)Y_{10}(\vec\Omega).
\end{equation}
An excitation of the form (\ref{e37}) thus 
radiates only an electric dipole contribution of the $(\ell=1,m=0)$ order.
All other multipoles remain unexcited.

It is interesting to compare our coherently excited initial state with 
an entangled single-excitation state of an extended medium of identical two-level
atoms
envisioned by Scully and collaborators \cite{SS091,SS092} as either having been prepared
by a swept-wave excitation or being intially in the symmetric Dicke state. 
While the entangled
atomic quantum states differ from our coherent initial excitation in essential ways,
they all share the important attribute of initial coherence,
as easily confirmed by the nonzero off-diagonal matrix elements of 
the density operator for the single-excitation state. We claim that it is
this spatially extended initial coherence, not entanglement {\it per se}, that is
fundamentally responsible for cooperative radiation processes such as 
superradiance and sub-radiance.
The absence of sub-radiant emission for the swept-excitation state
is a result of a coherent phasing of the emitted photon that yields
an enhanced emission rate in the forward direction. 
Both for our problem and the symmetric Dicke state, however, modes 
in which the atoms cooperate to trap coherent excitation and 
release it only weakly are also possible when $\beta >> 1$. 
But unlike the single-excitation state analysis, which is
based on a scalar-field approximation \cite{SS091,M09}, our exact
vector-field treatment accounts fully for the polarization and angular
distribution of the emitted radiation.

\subsection{\label{A.}Solution Procedure}

We first develop a general approach to the electric multipole radiation problem, and then
apply it to the special case of elctric dipole radiation we have just discussed.
Instead of formulating the problem in terms of an integral equation, as we
did for the MM radiation, we begin with the differential 
equations that describe the radiation fields inside and outside the 
spherical medium. In the rapid-transit approximation, which we
use, the retardation of the slowly varying amplitudes is negligible,
and that reduces Eq.~(\ref{e20}) to the simple form
\begin{equation}
\label{e39}
\cE(\vrt)=k_0^2\int \cP\left(
\vr^\prime,t\right){e^{ik_0|\vr-\vrp|}\over 
4\pi|\vr-\vrp|}d\vrp.
\end{equation}
The integral equation (\ref{e39}) can be turned into a differential equation
by operating on both sides of it with the operator $(\lap +k_0^2)$,
\begin{equation}
\label{e40}
(\lap +k_0^2)\cE(\vrt)=-k_0^2\cP(\vrt).
\end{equation}

We look for solutions of Eqs. (\ref{e21}) and (\ref{e40}) within the medium
that are expansions in spherical harmonics of the form
\begin{align}
\label{e41}
\cP^<(\vrt)=&\sumlm \cP^<_{\ell m}(r,t) Y_{\ell m}(\vec\Omega),\nonumber\\
\cE^<(\vrt)=&\sumlm \cE^<_{\ell m}(r,t) Y_{\ell m}(\vec\Omega).
\end{align}
The superscript $<$ on the variables $\cP$ and $\cE$ is here used to identify 
the polarization and field inside the medium. 
On writing the Laplacian as 
\begin{equation}
\label{e42}
\lap =\ddrsq +{2\over r}\ddr-{L^2\over r^2}
\end{equation}
and noting that the spherical harmonics $Y_{\ell m}$ are mutually orthogonal
eigenfunctions of the 
$L^2$ operator with eigenvalues $\ell(\ell+1)$, Eq.~(\ref{e40}) separates into 
individual equations, one for each spherical harmonic,
\begin{align}
\label{e43}
\left\{\ddrsq +{2\over r}\ddr+\left[k_0^2-{\ell(\ell+1)\over r^2}\right]
\right\}&\cE_{\ell m}^<(r,t)\nonumber\\
&=-k_0^2 \cP_{\ell m}^<(r,t).
\end{align}
The evolution of the polarization within  the medium, described by
Eq.~(\ref{e21}), also separates similarly,
\begin{equation}
\label{e44}
\ddt \cP_{\ell m}^<(r,t)=i{n_0^2\mus \over 3\hbar}\cE_{\ell m}^<(r,t).
\end{equation}

Outside the medium, {\it i.e.,} for $r>R$, the polarization vanishes 
identically, and the EM field obeys the free-space wave
equation obtained by setting $\cP_{\ell m}^<$ equal to 0 in Eq.~(\ref{e43}). Denoting 
the exterior field by the superscript $>$, we may thus write for $r>R$
\begin{equation}
\label{e45}
\left\{\ddrsq +{2\over r}\ddr+\left[k_0^2-{\ell(\ell+1)\over r^2}\right]
\right\}\cE_{\ell m}^>(r,t)=0.
\end{equation}
Our radiation problem thus separates both inside and outside the medium
into radiation by the individual multipoles.

The interior field and polarization are coupled to the exterior field by
the continuity of the normal component of the displacement field $\vec 
D_{\ell m}$ and of the tangential components of the electric field 
$\vE_{\ell m}$ at the boundary, $r=R$,
\begin{equation}
\label{e46}
\vr\cdot \vDllml|_{r=R}= \vr\cdot \vDglml|_{r=R}
\end{equation}
and
\begin{equation}
\label{e47}
\vr\times \vElml^< |_{r=R}= \vr\times \vElml^>
|_{r=R}.
\end{equation}

\subsection{\label{B.}The Eigenvalue Problem}

Because of the separability of the space and time variables in Eqs.~(\ref{e43})-(\ref{e45}),
they admit exponentially time-dependent solutions that maintain their shape
and are of the form
\begin{equation}
\label{e48}
\cE_{\ell m}(r,t)=\cE_{\ell m\lambda}(r) e^{-\lambda t}.
\end{equation}
For such solutions, Eq.~(44) defines the constitutive relation between the 
polarization and the field
\begin{equation}
\label{e49}
\cPlml=-i{n_0\mus\over 3\hbar\lambda}\cElml^<(r).
\end{equation}
When combined with Eq.~(\ref{e43}), this relation leads to the 
equation for $\cElml^<$:
\begin{equation}
\label{e50}
\left\{\ddrsq +{2\over r}\ddr+\left[\gl^2-{\ell(\ell+1)\over r^2}\right]
\right\}\cE_{\ell m\lambda}^<(r)=0.
\end{equation}

A general solution of Eq.~(\ref{e50}) that is also finite at $r=0$ must, as we have 
seen earlier, take the form
\begin{equation}
\label{e51}
\cElml^<(r)=Aj_\ell(\gl r)
\end{equation}
within the medium.
Outside the medium, the fields consist of purely radiative, out-going waves. 
They are described by a solution of Eq.~(\ref{e45}) of form
\begin{equation}
\label{e52}
\cElml^>(r)=Bh^{(1)}_\ell(k_0 r).
\end{equation}
From these field forms, we obtain, as we now show, the 
electric, magnetic, and displacement fields throughout space.

Returning to the definition of $\eta$ in Eq.~(\ref{e15}), of which $\cE$ is the 
slowly varying envelope, we note that the full electric field $\vec E_{\ell m\lambda}$ and the radial component $\cElml Y_{\ell m}$ of its envelope are 
related as follows:
\begin{equation}
\label{e53}
\vec L\cdot \grad\times \vec E_{\ell m\lambda}=\cElml(r) Y_{\ell m} (\vOm) e^{-\lambda t} e^{-i\omega_0 t}.
\end{equation}
From Faraday's law, which, in the slowly-varying envelope approximation, 
takes the form, $\grad\times \vec E_{\ell m\lambda}\approx ik_0 \vec B_{\ell
m\lambda}$, we have
$$\vec L\cdot \vec B_{\ell m\lambda}={1\over ik_0} \cElml(r) Y_{\ell m}(\vOm) 
e^{-\lambda t} e^{-i\omega_0 t},$$
an equation that has the simple solution \cite{Jackson1}
\begin{equation}
\label{e54}
\vec B_{\ell m\lambda} =
{1\over ik_0\ell(\ell+1)} \cElml(r) \vec L Y_{\ell m}(\vOm) 
e^{-\lambda t} e^{-i\omega_0 t},
\end{equation}
consistent with the requirement that $\vec B$ has no radial component for 
pure EM radiation.
If we now use the Maxwell-Amp\`ere law in which the time derivative $(\partial
\vec D/\partial t)$ is approximately set equal to $-i\omega_0 \vec D$, we
obtain the corresponding displacement vector
\begin{align}
\label{e55}
\vec D_{\ell m\lambda}&\approx -{1\over ik_0}\grad\times \vBlml\nonumber\\
&={1\over k_0^2
\ell(\ell+1)}\left[\grad\times \cElml(r) \vec L Y_{\ell m}(\vOm)\right]
e^{-(\lambda+i\omega_0) t}.
\end{align}
Equation (\ref{e55}) also describes the electric field in the free space
outside the medium.
Finally, the electric field $\vec E^<_{\ell m\lambda}$ within the medium
is obtained by means of Eq.~(\ref{e49}), which yields
$$\vec E^<_{\ell m\lambda}=\vec D^<_{\ell m\lambda}-\vec P_{\ell m\lambda}
= \vec D^<_{\ell m\lambda}+{in_0\mus\over 3\hbar \lambda}\vec E^<_{\ell m
\lambda}.$$
This equation is easily solved for $\vec E^<_{\ell m\lambda}$,
\begin{equation}
\label{e56}
\vec E^<_{\ell m\lambda}={k_0^2\over \gl^2}\vec D^<_{\ell m\lambda},
\end{equation}
an expression in which use has been made of the definition (\ref{e33}) for $\gl$.
The ratio $(\gl/k_0)^2$, the electrical permittivity of the medium for
the propagation of the $(\ell m\lambda)$ mode, naturally connects the 
electric and displacement fields within the medium.

We are now in a position to apply our boundary conditions (\ref{e46}) and (\ref{e47}). Taking the
scalar product of Eq.~(\ref{e55}) with $\vr$ and noting that 
$$\vr\cdot (\grad\times \vec F)=(\vr\times \grad)\cdot \vec F=i\vec L\cdot 
\vec F$$
for an arbitrary vector field $\vec F$ and that $\vec L$ commutes with 
any function of the radial coordinate $r$, we find the relation
\begin{equation}
\label{e57}
\vec r\cdot \vec D_{\ell m\lambda}={i\over k_0^2}\cElml(r)\left[{L^2 
Y_{\ell m}(\vOm)\over \ell(\ell+1)}\right]
={i\over k_0^2}\cElml(r) Y_{\ell m}(\vOm).
\end{equation}
In view of this relation, the condition (\ref{e46}) amounts simply to the
continuity, at the boundary $r=R$, of the electric field amplitude 
$\cElml$ given by Eqs.~(\ref{e51}) and (\ref{e52}), 
\begin{equation}
\label{e58}
Aj_\ell (\gl R)=Bh^{(1)}_\ell(k_0 R).
\end{equation}

The expression for the tangential components of $\vec E_{\ell m\lambda}$
is more involved. By taking the vector product of Eq.~(\ref{e55}) with 
$\vr$ and using the rule $\vec A\times(\vec B\times\vec C)=(\vec A\cdot
\vec C)\vec B-(\vec A\cdot \vec B)\vec C$ while allowing for the correct
spatial derivatives to be taken, we have
\begin{eqnarray*}
\vr\times \vec D_{\ell m\lambda}=
{1\over k_0^2
\ell(\ell+1)}&&\left[r_i\grad \cElml(r) L_i Y_{\ell m}\right. \\
&-&r\left. \ddr\cElml(r) \vec L 
Y_{\ell m}\right],
\end{eqnarray*}
in which it is understood that the repeated vector-component index $i$ 
is summed over its three values.
Noting now that $f\grad g=\grad (fg)-g\grad f$ and that $r_i L_i=\vr\cdot
\vec L=0$, we have the expression for $\vr\times \vec D_{\ell m\lambda}$
in terms of the amplitude $\cElml$
\begin{equation}
\label{e59}
\vr\times \vec D_{\ell m\lambda}=
-{1\over k_0^2
\ell(\ell+1)}\left[\cElml+r \cElml^\prime(r)\right] \vec L Y_{\ell m}.
\end{equation}
But since $\vec E^>_{\ell m\lambda}=\vec D^>_{\ell m\lambda}$ and from
Eq.~(\ref{e56})
$\vec E^<_{\ell m\lambda}={k_0^2\over \gl^2}\vec D^<_{\ell m\lambda}$,
the boundary condition (\ref{e47}) may be written as the matching condition
\begin{equation}
\label{e60}
{1\over \gl^2}\left[\cElml^<+R\cElml^{<\,\prime}(R)\right]
={1\over k_0^2}\left[\cElml^>+R\cElml^{>\,\prime}(R)\right].
\end{equation}
When Eqs.~(\ref{e51}) and (\ref{e52}) are substituted into this relation, we
secure the second condition that our solutions must obey,
\begin{align}
\label{e61}
{A\over \gl^2}\left[ j_\ell(\gl R)\right. &+\left.\gl R j_\ell^\prime (\gl R)\right]\nonumber\\
&={B\over k_0^2}\left[ h_\ell^{(1)}(k_0 R)+k_0 R h_\ell^{(1)\,\prime} (k_0 R)
\right].
\end{align}

The two conditions (\ref{e58}) and (\ref{e61}) must be met simultaneously, and we have, by
taking their ratio, the eigenvalue equation
\begin{equation}
\label{e62}
{(xj_\ell(x))^\prime\over x^2j_\ell(x)}
={(\beta h_\ell^{(1)}(\beta))^\prime\over \beta^2
h_\ell^{(1)}(\beta)},
\end{equation}
where 
\begin{equation}
\label{e63}
\beta=k_0R\ \ {\rm and} \ \ x\equiv\gl R=\beta\sqrt{1-i{n_0\mus\over 3\hbar\lambda}} .
\end{equation}
This eigenvalue relation applies to radiation from an electric multipole 
of arbitrary order $(\ell,m)$. For our special initial condition (\ref{e37}), we 
need, however, only consider the azimuthally symmetric, electric dipole
radiation with $\ell=1, m=0$.

\subsubsection{\label{(i)}Electric Dipole Radiation}

Since 
$$ (xj_1(x))^\prime =\sin x+{\cos x\over x}-{\sin x\over x^2}$$
and
\begin{equation}
\label{e64}
(\beta h^{(1)}_1(\beta))^\prime =
\left(-i +{1\over \beta}+{i\over \beta^2}\right)e^{i\beta},
\end{equation}
Eq.~(\ref{e62}) reduces to the form
$${\sin x\over \sin x-x\cos x}-{1\over x^2}={i\over \beta}-{i\over \beta
(\beta^2+i\beta)},$$
from which, by a simple transposition and division of both sides of
the equation by $\sin x$, the following transcendental form for the 
eigenvalue condition results:
\begin{equation}
\label{e65}
x \cot x=1-{\beta x^2\over \beta+ix^2\left(1-{\displaystyle 1\over 
\beta^2+i\beta} \right)}.
\end{equation}
Although still complicated in appearance, the form (\ref{e65}) for the eigenvalue
equation permits an asymptotic analysis of the roots $x=\gl R$ -- and hence of the 
eigenvalues $\lambda$ -- in the limits of large and small radius,
$\beta>>1$ and $\beta<<1$.

\subsubsection{\label{ii}Electric Dipole Radiation from Small Spheres, $\beta<<1$}

The superradiant mode that we discussed in Sec.~II is but one of
a sequence of exponentially decaying radiation modes that one can discuss for a
small sphere. As we shall see presently, all of the other modes, however,
radiate quite weakly by comparison. 

The superradiant mode has a propagation constant $x=\gl R$ that is of order $\beta$,
as we shall see shortly. For $\beta<<1$, we may expand the left hand side
of the eigenvalue equation (\ref{e65}) in powers of $\beta$, retaining only
terms of the most significant order separately in the real and imaginary parts
on the left hand side to obtain the approximate equation
$$x\cot x\approx {1\over (1-x^2/\beta^2)}-i{x^2\beta\over (1-x^2/\beta^2)^2}.$$
If we now expand the left hand side in powers of $x$ and keep only the
two most significant terms, we may further approximate the eigenvalue relation
for small $|x|$ as
$$x^2\approx -2\beta^2+i{3x^2\beta^3\over (1-x^2/\beta^2)},$$
an equation that may be solved by an iterative procedure. To the lowest
significant order in its real and imaginary parts, the solution
assumes the expression
\begin{equation}
\label{e66}
x^2\approx -2\beta^2-2i\beta^3.
\end{equation}
By employing the relation (\ref{e63}) between $x$ and the eigenvalue $\lambda$, we 
then obtain 
\begin{equation}
\label{e67}
\lambda\approx {2n_0\mus\beta^3\over 27\hbar}+i{n_0\mus\over 9\hbar}.
\end{equation}
This expression for the complex decay constant of the fundamental
mode coincides with that obtained in Sec.~II by means of physical considerations. 

For all other eigenmodes, $x$ is of order 1 or larger, so
we can neglect the term $\beta$ in comparison with
$ix^2$ in the denominator
of the right hand side of the eigenvalue equation (\ref{e65}) and obtain the
approximate expression,
$$x\cot x=1-i{\beta^2(i+\beta)\over 1-i\beta-\beta^2}.$$
By expanding the denominator for small $\beta$ and keeping only terms
of the most significant order separately in the real and imaginary parts of 
the resulting right hand side of the preceding equation,
we may approximate it as
\begin{equation}
\label{e68}
x\cot x=1+i\beta^5.
\end{equation}
Solutions of Eq.~(\ref{e68}) of form 
\begin{equation}
\label{e69}
x_n=\left(n-{1\over 2}\right)\pi +\epsilon_n, \ \ \ \ \ n=1,2,...,
\end{equation}
where $|\epsilon_n|<<1$, may be found, as we see by substituting Eq.~(\ref{e69})
into Eq.~(\ref{e68}). To the lowest order in $\epsilon_n$,
the following expression for the allowed values of $x$ thus results:
\begin{equation}
\label{e70}
x_n=\left(n-{1\over 2}\right)\pi -i{\beta^5\over
\left(n-{1\over 2}\right)\pi}.
\end{equation}
From the relation (\ref{e63}) between $x$ and $\lambda$ for any mode,
we may now solve for the latter. On keeping
only terms to the lowest order in $\beta$ separately for the real
and imaginary parts, we obtain the $n$th eigenvalue
\begin{align}
\label{e71}
\lambda_n=&{2\over 3}{n_0\mus\over \hbar}{\beta^5\over (n-1/2)^4\pi^4}\nonumber\\
&-i {n_0\mus\over 3\hbar}{\beta^2\over (n-1/2)^2\pi^2}, \ \ \ n=1,2,\ldots
\end{align}

Both the frequency shift and decay rate for each of these modes are 
suppressed by a factor O$(\beta^2)$ relative to those for the superradiant 
mode. The dramatic decrease of the decay rates with increasing 
mode index shows that the excitations of these modes remain trapped for long periods.

Even when $\beta$ is of order 1, the behavior of the eigenvalues, $\lambda_n$, is qualitatively similar
to that for $\beta<<1$ we have just considered. We demonstrate this by
plotting the real and imaginary parts of the first 20 eigenvalues for $\beta=1$ in Figs.~1(a) and 1(b). 
The existence of only a single superradiant mode is quite clear. The next four most strongly decaying modes
have decay rates that are roughly 50, 500, 2000, and 6000 times smaller.
Their frequency detunings from atomic resonance, which are of the opposite sign relative to 
that for the single superradiant mode, decrease more modestly, however, so their successive differences
are quite small. These features of the weakly decaying modes are responsible, as we shall see later,  
for a slow oscillatory decay of efficiently trapped excitation from a small sphere which 
is initially excited from its center out to only a small fraction of its total radius.

\begin{figure}[ht]
     \centering
     \subfigure[]
     {\label{fig:DecayRates1}
     \includegraphics[width=.4\textwidth, height=0.25\textheight]{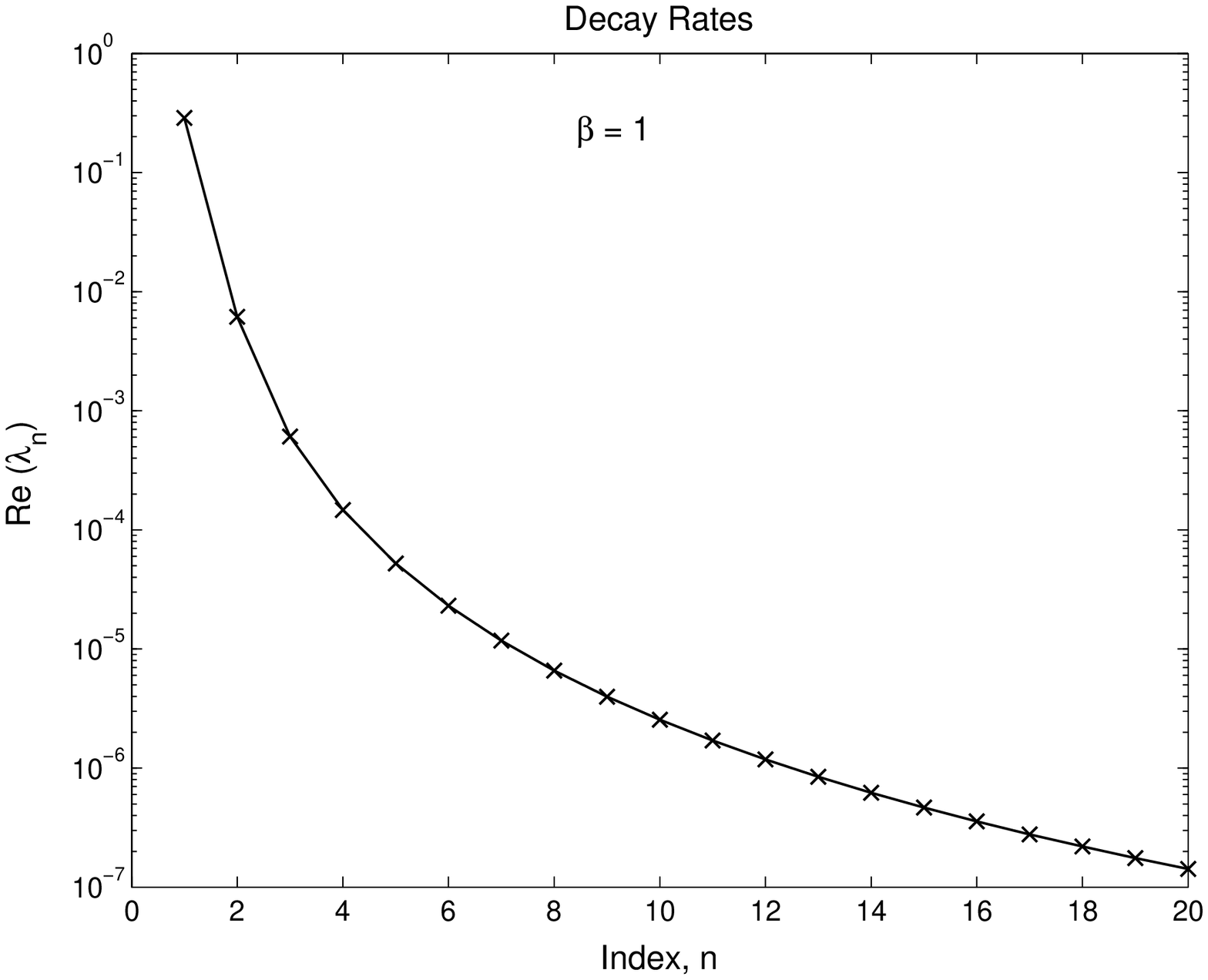}}

     \subfigure[]
     {\label{fig:FreqShifts1}
     \includegraphics[width=.4\textwidth, height=0.25\textheight]{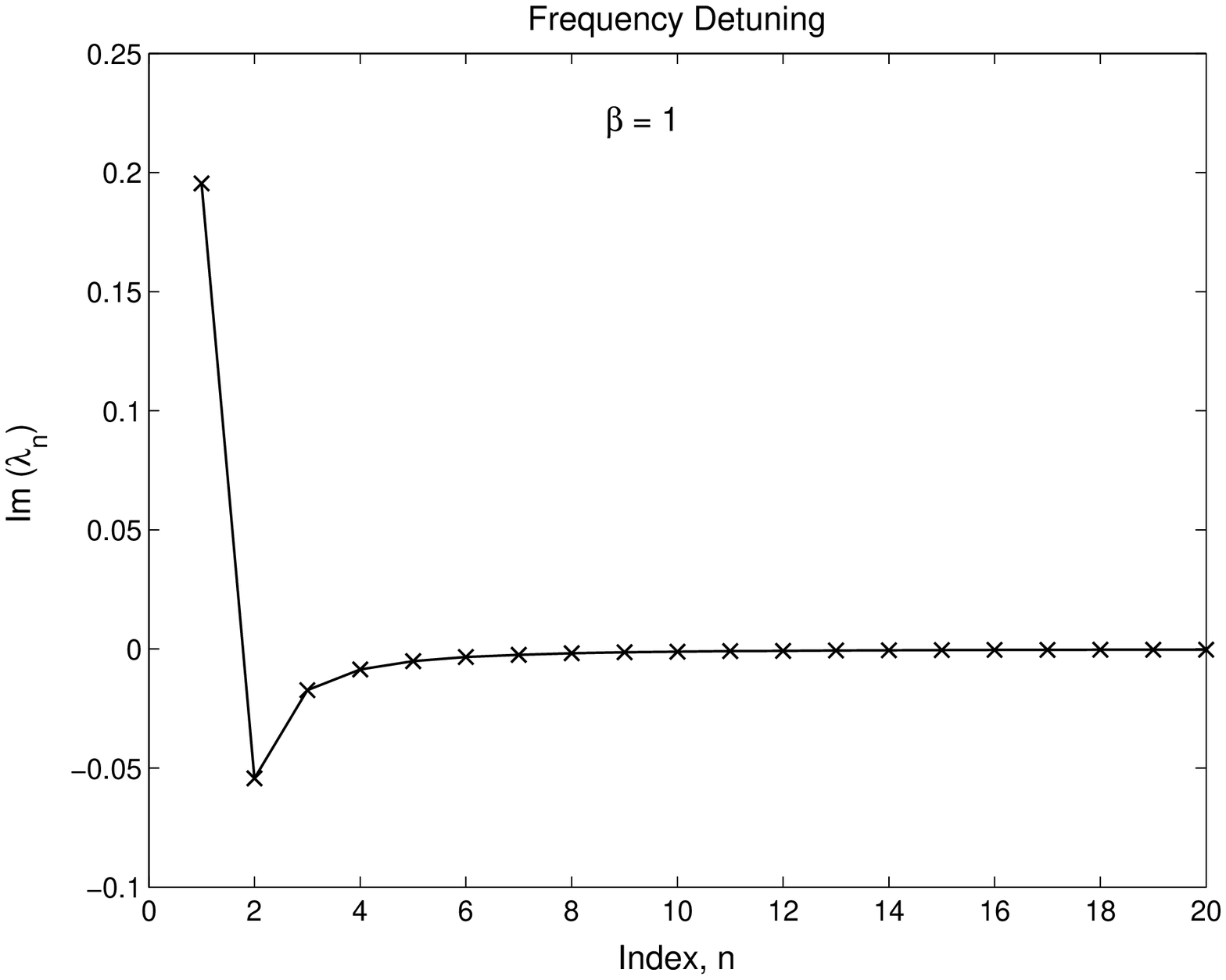}}
     \caption{(a) Decay rates and (b) frequency detuning (from resonance) for the 20 fastest decaying
                  modes, for $\beta=1$, in units of $n_0\mu^2\beta/(3\hbar)$.}
\end{figure}

\subsubsection{\label{(iii)}Electric Dipole Radiation from Large
Spheres, $\beta>>1$}

When the radius of the sphere greatly exceeds the wavelength, 
$\beta>>1$, we may ignore $(\beta^2+i\beta)^{-1}$ when compared to 1,
and Eq.~(\ref{e65}) simplifies somewhat to the expression
\begin{equation}
\label{e72}
x\cot x=1-
{\beta x^2\over \beta+ix^2} .
\end{equation}
If the inequality $|x|^2>>\beta$ also holds, then
the eigenvalue equation simplifies still further to the form
\begin{equation}
\label{e73}
x\cot x\approx 1+i\beta\approx i\beta.
\end{equation}
This equation is formally identical to the eigenvalue equation for the odd
modes in the one-dimensional problem of resonant propagation inside a
slab that we discussed in I, provided the slab thickness $L$ is taken to be
the sphere diameter $2R$. This 
isomorphism between the spherical and slab geometries in the limit
of large medium extensions holds only for those modes that have propagation 
constants $\gl$ that in magnitude greatly exceed the curvature $1/R$ of the 
spherical surface, consistent with the mathematical requirements $|x|^2>>
\beta$ and $\beta>>1$.

Because of the infinitely many branches of the cotangent function, the eigenvalue 
equation (\ref{e72}) has infinitely many solutions. However, only those solutions for which
$|x|$ is comparable to $\beta$ correspond to field eigenmodes that, being spatially 
phase matched to the radiation field, are strongly coupled to it, and thus radiate 
efficiently. The eigenvalues of the other, relatively weakly radiating modes are
analogous to those we have already considered for the one-dimensional problem in I, and
need no further attention.

When $|x|\sim \beta$, then $|x^2|\sim\beta^2>>\beta$ and the right hand side of
Eq.~(\ref{e72}) may be approximated to the lowest order in its real and imaginary parts by the
expression $i\beta+(1-\beta^2/x^2)$. Now dividing both sides of Eq.~(\ref{e72}) by $x\sim\beta$,
we may reduce it to the following approximate form:
\begin{equation}
\label{e74}
\cot x\approx i+{2\over \beta}\left(1-{\beta\over x}\right),
\end{equation}
where we have only retained the most significant power of the small quantity
$(1-\beta/x)$ in the real and imaginary parts separately. By substituting $x=-iy$ and
$\cot x=i\coth y$ in Eq.~(\ref{e74}), we see that $\coth y$ differs little from 1, which
implies that $y$ must have a large real part and $\coth y\approx 1+2\exp(-2y)$.
With this approximation, Eq.~(\ref{e74}) can be solved for its multiple roots. They may
each be labeled by an integer $n$, with the value
\begin{equation}
\label{e75}
x_n=-iy_n\approx (n+1/4)\pi-(i/2)\ln \beta.
\end{equation}
For consistency with the assumption $|x|\sim\beta$, we require that $n\pi\sim \beta$.
The eigenvalues $\lambda_n$ now follow from their relation (\ref{e63}) with $x_n$,
\begin{eqnarray}
\label{e76}
\lambda_n&\approx& {in_0\mus/(3\hbar) \over 1-[(n+1/4)^2\pi^2/\beta^2]+i[(n+1/4)\pi/
\beta^2]\ln\beta}\nonumber\\
&\approx& {in_0\mus\beta/(6\hbar)\over (\beta-n\pi)+i(1/2)\ln\beta},
\end{eqnarray}
where the last approximate equality follows from dropping the 1/4 when compared to
$n$, and then setting $n\pi\approx \beta$ and $(\beta^2-n^2\pi^2)\approx 2(\beta-n\pi)$. 
The real and imaginary parts of the eigenvalues $\lambda_n$ thus have  
a simple resonant character, similar in form to the frequency
dependence of the imaginary and real parts, respectively, of the susceptibility 
of a resonant dielectric medium.

In Figs. 2(a) and 2(b), we display for $\beta=1000$ the real and imaginary parts of 
the eigenvalues in the resonant domain, $n\pi\sim\beta$. These have been obtained
both by a highly accurate numerical treatment of the exact eigenvalue equation (\ref{e65}) 
and by means of the analytical approximation (\ref{e76}). The analytical approximation is evidently 
quite accurate for such a large value of $\beta$.

\begin{figure}[ht]
     \centering
     \subfigure[]
     {\label{fig:DecayRates1000}
     \includegraphics[width=.4\textwidth, height=0.25\textheight]{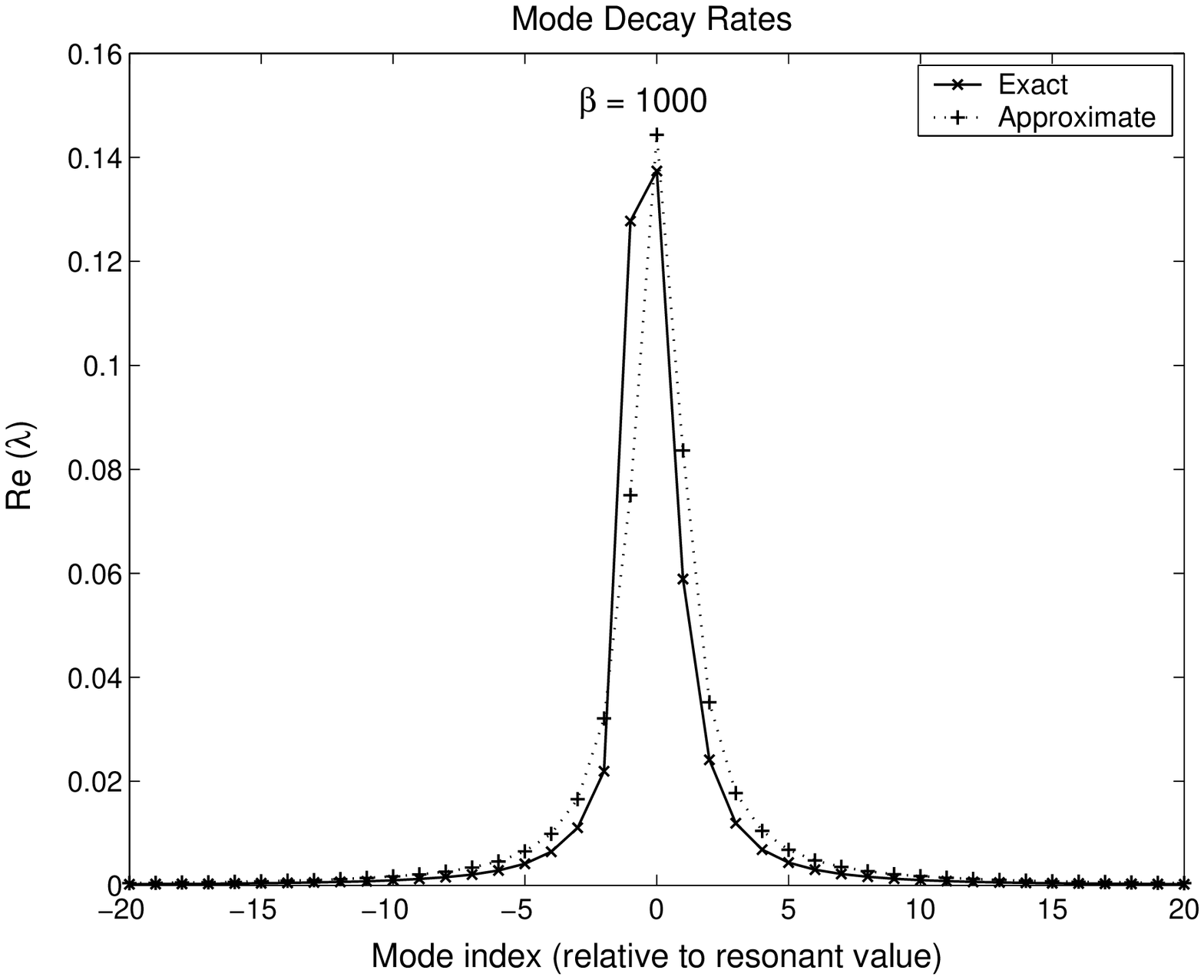}}

     \subfigure[]
     {\label{fig:FreqShifts1000}
     \includegraphics[width=.4\textwidth, height=0.25\textheight]{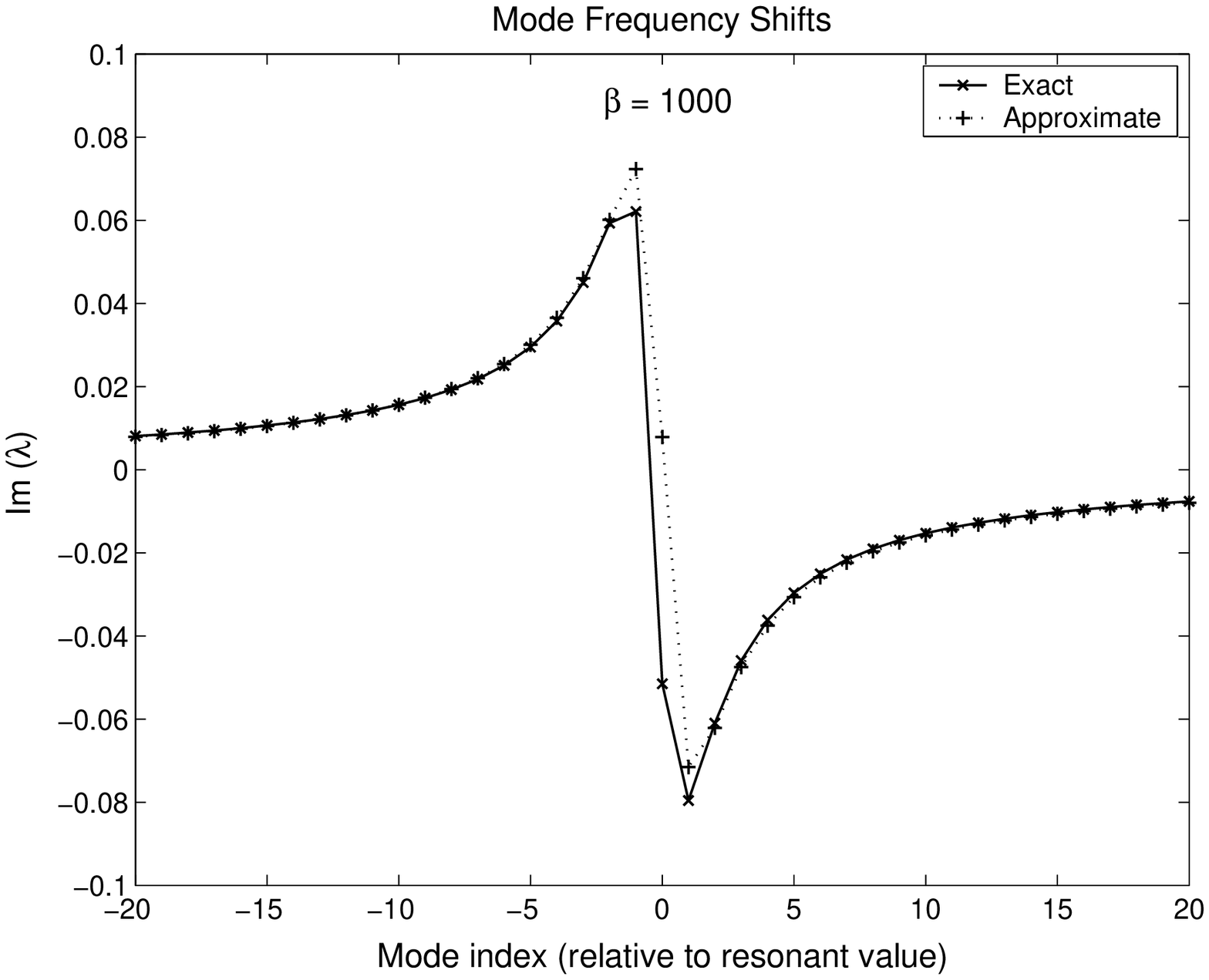}}
     \caption{(a) Decay rates and (b) frequency detuning (from resonance) for the 40 fastest decaying
                  modes, for $\beta=1000$, in units of $n_0\mu^2\beta/(3\hbar)$.}
\end{figure}

For further discussions of EM radiation from large spheres we undertake the development of 
the orthogonality properties of the EM eigenmodes. They are somewhat less self-evident than the 
analogous properties of the MM eigenmodes we considered in Sec.~IV.  

\section{\label{VI}Orthogonality Properties of the Electric Multipole Modes}

For the $n$th mode of the $(\ell m)$ order, which we shall call simply
the $(\ell mn)$ mode, the magnetic field $\vBEMlmn$ 
is given by Eq.~(\ref{e54}), and the electric field $\vEEMlmn$ essentially by its curl,
\begin{equation}
\label{e77}
\vBEMlmn={1\over ik_0\ell(\ell+1)}\cElmn(r) \vec L Y_{\ell m}
(\vOm),
\end{equation}
\begin{equation}
\label{e78}
\vEEMlmn={ik_0\over \gln^2}\grad\times \vBEMlmn.
\end{equation}
Let us first consider the orthogonality integral for the electric fields 
of two modes $p,p^\prime$ in two different multipole orders
$(\ell,m)$ and $(\ell^\prime,m^\prime)$, with $\ell\neq\ell^\prime$:
$$\int_{r<R} \vEEMlmn\cdot\vEEMlpmpnp d\vr.$$
Use of Eq.~(\ref{e78}) to replace the electric field $\vEEMlmn$ in terms of 
the corresponding magnetic field $\vBEMlmn$ and the identity 
\begin{equation}
\label{e79}
\grad\cdot(\vec A\times\vec B)=\vec B\cdot \grad\times\vec A-\vec A\cdot\grad
\times \vec B,
\end{equation}
followed by the use of Gauss's theorem to reduce a total divergence term to
a surface integral over the sphere, yields the result
\begin{align}
\label{e80}
\int_{r<R} &\vEEMlmn\cdot\vEEMlpmpnp d\vr\nonumber\\
&={ik_0\over\gln^2}\int_{r=R}\vBEMlmn\times
\vEEMlpmpnp \cdot\hat r d^2S \nonumber\\
&-{k_0^2\over\gln^2}\int_{r<R}\vBEMlmn\cdot\vBEMlpmpnp d\vr.
\end{align}
Here Faraday's law was employed in the last term to express the curl of 
$\vEEMlpmpnp$ in terms of
the magnetic field $\vBEMlpmpnp$ of the $(\ell^\prime m^\prime n^\prime)$ mode.
That the last term in Eq.~(\ref{e80}) vanishes unless $\ell= \ell^\prime$ and
$m=-m^\prime$ follows from
the orthogonality of the different vector spherical harmonics $\vec L Y_{\ell m}$
in terms of which the magnetic fields of the two modes may be expressed by means of Eq.
(\ref{e77}). By a similar but more detailed argument involving
the identity \cite{Jackson1} 
\begin{displaymath}
i\vec\nabla\times\vec L = \vec r \nabla^2-\vec\nabla\left(1+r{\partial\over\partial r}\right),
\end{displaymath}
we may reduce the surface integral in Eq.~(\ref{e80}) to a form involving 
$\int d^2\Omega \, \vec L Y_{\ell m}\cdot \vec LY_{\ell^\prime,m^\prime}$,
which too vanishes unless $\ell=\ell^\prime$ and $m=-m^\prime$.
This proves the orthogonality of the vector electric fields for modes with different
$\ell$ and $m$ values.

A different approach is needed to establish orthogonality rules for modes with the 
same $\ell$ value. We begin with the field equations that $\vEEMlmn$ and $\vEEMlmpnp$
obey,
\begin{equation}
\label{e81}
-ik_0\grad\times\vBEMlmn+\gln^2\vEEMlmn=0,
\end{equation}
\begin{equation}
\label{e82}
-ik_0\grad\times\vBEMlmpnp+\glnp^2\vEEMlmpnp=0.
\end{equation}
Taking the scalar product of Eq.~(\ref{e81}) with $\vEEMlmpnp$ and subtracting the result
from that obtained on taking the scalar product of Eq.~(\ref{e82}) with $\vEEMlmn$,
and integrating the difference over the spherical sample we secure the result
\begin{align}
\label{e83}
&(\gln^2-\glnp^2)\int_{r<R} \vEEMlmn\cdot\vEEMlmpnp d\vr\nonumber\\
&=ik_0\int_{r<R}
\left[\vEEMlmpnp \cdot\grad\times\vBEMlmn\right.\nonumber\\
&\qquad\qquad -\left. \vEEMlmn \cdot\grad\times\vBEMlmpnp\right]d\vr.
\end{align}
We now use the identity (\ref{e79}) to reexpress each term on the right hand side of Eq.~(\ref{e83})
in terms of a complete divergence, which reduces to a surface integral according to
Gauss's law, and a curl term, which from Faraday's law can be written
as a volume integral of $\vBEMlmn\cdot\vBEMlmpnp$. We see in this way that
Eq.~(\ref{e83}) simplifies to a single surface integral, the two volume integrals canceling
each other out,
\begin{align}
\label{e84}
&(\gln^2-\glnp^2)\int_{r<R} \vEEMlmn\cdot\vEEMlmpnp d\vr\nonumber\\
&=ik_0\int_{r=R}
\left[(\hat r\times\vEEMlmn)\cdot \vBEMlmpnp\right.\nonumber\\
&\qquad\qquad-\left.
(\hat r\times\vEEMlmpnp)\cdot \vBEMlmn\right]d^2S.
\end{align}
The surface integral involves only the tangential components of 
both the electric and magnetic fields, which are all continuous across the
spherical boundary. The surface integral can, as such, be expressed equally 
well in terms of the same components of the free-space fields in the immediate
exterior of the sphere. However, since in each multipole order $(\ell m)$
the exterior fields are all expressed in terms of the outgoing wave solutions
$h_\ell^{(1)}(k_0r) Y_{\ell m}(\vOm)$ and their derivatives, independent of
the mode indices $n,n^\prime$, the surface integral on the right hand side of
Eq.~(\ref{e84}) vanishes identically. It follows then from Eq.~(\ref{e84}) that if $\gln
\neq\glnp$, then the orthogonality integral vanishes,
\begin{equation}
\label{e85}
\int_{r<R} \vEEMlmn\cdot\vEEMlmpnp d\vr\sim\delta_{nn^\prime}.
\end{equation}

Combining all of the specific orthogonality relations we have discussed
so far in this section, we may write down the overall orthogonality relation
\begin{equation}
\label{e86}
\int_{r<R} \vEEMlmn\cdot\vEEMlpmpnp d\vr\sim
\delta_{\ell\ell^\prime} \delta_{m,-m^\prime} \delta_{nn^\prime}.
\end{equation}
This orthogonality relation (\ref{e86})
may also be directly established by using Eq.~(\ref{e77}) and (\ref{e78}) to express its left 
hand side in terms of the mode functions $\cElmn\sim j_\ell(\gln r)$, simplifying
the resulting integral by means of angular momentum operator identities, and
then exploiting the eigenvalue relation (\ref{e62}) that each mode must obey. 

It is worth noting that the requirement $m=-m^\prime$, rather than
$m=m^\prime$, for the nonvanishing
of the expression (\ref{e86}) is a reminder of the symmetric but non-Hermitian character of the propagation
kernel for the electromagnetic field. This nonhermiticity was already noted in II in the
context of one-dimensional propagation when an incident wave with transverse magnetic
polarization was obliquely incident on a slab.

\subsection{\label{A}Expansion of an Arbitrary EM Field in the Corresponding Modes}

The orthogonality integral (\ref{e86}) makes it possible to expand an arbitrary electromagnetic
field in a particular EM order $(\ell, m)$ in terms of the EM modes, given by Eqs.~(\ref{e77}) 
and (\ref{e78}), of the same order. 
Let $\grad\times[f(r)\vec L Y_{\ell m}]$ be an arbitrary field in this
order expressed as the mode sum 
\begin{equation}
\label{e87}
\grad\times\left[f(r)\vec L Y_{\ell m}\right]=\sum_{n=1}^\infty f_n 
\grad\times\left[j_\ell(\gln r)\vec L Y_{\ell m}\right].
\end{equation}
The coefficients $f_n$ are obtained by multiplying both sides of Eq.~(\ref{e87}) by the mode
function $\grad\times[j_\ell(\glnp r)\vec L Y_{\ell, -m}]$, integrating over the spherical
sample, and utilizing the orthogonality relation (\ref{e86}),
\begin{equation}
\label{e88}
f_n = {(j_\ell(\gln r),f(r))\over (j_\ell(\gln r),j_\ell(\gln r))},
\end{equation}
where the symbol $(f,g)$ defines an inner product, 
\begin{equation}
\label{e89}
(f,g)\equiv \int_{r<R}d\vec r \left\{\grad\times [f(r)\vec L Y_{\ell,-m}]
\right\}\cdot \left\{\grad\times [g(r)\vec L Y_{\ell m}]\right\}. 
\end{equation}
Whenever either $f$ or $g$ is a mode amplitude function $j_\ell(\gln r)$, the 
inner product (\ref{e89}) can be simplified greatly. To see this, 
we make use of formula (\ref{e79}) in Eq.~(\ref{e89}) to transform its integrand 
to a term involving $\grad\times(\grad\times[g(r)\vec L Y_{\ell m}])$, which, when $g$ is
a mode function, is simply $\gln^2 g(r)\vec L\,Y_{\ell m}$, along with a pure divergence term, a
term that from Gauss's theorem is equal to a surface integral. Thus, when $g$ is a mode
function, Eq.~(\ref{e89}) takes a simpler form,
\begin{align*}
(f,g)=\gln^2&\int_{r<R}f(r)g(r)\vec LY_{\ell,-m}\cdot\vec LY_{\ell m} d\vr\\
+&
\int f(r)\vec LY_{\ell,-m}\times[\grad\times(g \vec LY_{\ell m})]\cdot\hat r\, d^2S.
\end{align*}
Use of the solid-angle integral identity
$$\int d^2\Omega \vec LY_{\ell,-m}\cdot\vec LY_{\ell m}=(-1)^{m+1}\ell(\ell+1)$$
and the fact \cite{Jackson1} that the surface integral in the preceding equation is 
simply $(-1)^{m+1}\ell(\ell+1)
Rf(R)(d/dR)[Rg(R)]$ reduces the inner product (\ref{e89}) to the form
\begin{align}
\label{e90}
(f,g)=(-1)^{m+1}&\ell(\ell+1)\left\{Rf(R){d\over dR}\left[Rg(R)\right]\right.\nonumber\\
&\left.+\gln^2\int_0^R
f(r) g(r) r^2 dr\right\}.
\end{align}
This form of the inner product can also be used as the starting point of a simple, direct
proof of the orthogonality of different modes in a given multipole order.

\section{\label{VII}Radiation from a Large Sphere with a Uniformly Excited Spherical Inner Core}

Coherent radiation from a large resonant sphere will, in general, 
involve a large number of modes of both multipole types and their infinitely many orders.
A particularly simple situation occurs, however, when 
an inner concentric spherical region of the sphere is uniformly excited initially, as shown in Fig.~3.
This is a special case of a polarization with radially symmetric amplitude and uniform 
direction, which emits pure electric dipole radiation in the $(1,0)$ order.

\begin{figure}[ht]
     \includegraphics[width=.4\textwidth, height=0.3\textheight]{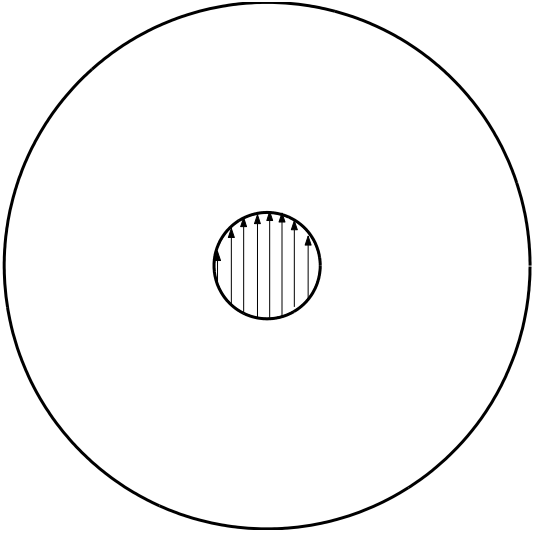}
       \caption{Spherical medium of radius $R$ with an excited concentric core of radius $r_0=fR$, $f\leq 1$.}
\end{figure}

Let us assume then an initial polarization of form
\begin{equation}
\label{e91}
\vP(\vr,0)=\begin{cases}
\hat z P_0 & {\rm for\ } r<r_0\\ 0 & {\rm for\ } r_0<r<R.
\end{cases}
\end{equation}
The electric field that this polarization initially radiates has a simple expression 
in the rapid transit approximation \cite{PG00} 
in which retardation of slowly varying amplitudes is ignored and the second time 
derivatives in Eq.~(\ref{e14}) are replaced by $-\omega^2$. In this approximation,
the initial electric field $\vEp(\vr,0)$ may be expressed 
as an integral of the scalar product of the tensor propagator 
\begin{equation}
\label{e92}
{\bf G}(\vr-\vrp)=\left({\bf 1}+{1\over k_0^2}\grad\grad\right)
{e^{ik_0|\vr-\vrp|}\over 4\pi |\vr-\vrp|}
\end{equation}
and the initial polarization (\ref{e91}) over the spherical sample, 
which reduces to the form
\begin{equation}
\label{e93}
\vE(\vr,0)\approx P_0\left(k_0^2\hat z+{\p\over \p z}\grad\right)\int_{r^\prime<r_0}
{e^{ik_0|\vr-\vrp|}\over 4\pi |\vr-\vrp|}d\vrp.
\end{equation}
The initial field amplitude $d(r,0)$ is the radial
function that multiplies the spherical harmonic $Y_{10}(\vOm)$ in the 
quantity $\vL \grad\times\vE(\vr,0)$. Taking the curl
of the left hand side of Eq.~(\ref{e93}) eliminates the pure gradient term,
\begin{equation}
\label{e94}
\grad\times \vE(\vr,0)=k_0^2P_0\grad \int_{r^\prime<r_0}
{e^{ik_0|\vr-\vrp|}\over 4\pi |\vr-\vrp|}d\vrp\times \hat z.
\end{equation}
By using the identity (\ref{e25}) to replace the integrand in Eq.~(\ref{e94}) by a spherical 
harmonic sum and then integrating it over the angular coordinates,
we see that the integral in Eq.~(\ref{e94}) is a purely radial function,
\begin{equation}
\label{e95}
f(r)=ik_0\int_0^{r_0} j_0(k_0 r^<) h_0^{(1)}(k_0r^>) r^{\prime 2}dr^\prime.
\end{equation}
We thus have the result
$$\grad\times \vE(\vr,0)=k_0^2P_0\grad f(r)\times \hat z={df\over dr}{\vec r\over r}\times 
\hat z. $$
Operating on both sides of this equation by $\vL$, noting that any function
of $r$ commutes with $\vec L$, and using the result
\begin{align*}
\vL (\vr\times \hat z)&={1\over i} (\vr\times \grad)\cdot(\vr\times \hat z)\nonumber\\
&={1\over i} [\vr\cdot(\hat z\cdot\grad)\vr-(\vr\cdot\hat z)(\grad\cdot\vr)]\\
&=2i\vr\cdot z=2ir\cos \theta=2i\sqrt{4\pi\over 3}rY_{10}(\vOm),
\end{align*}
we obtain 
\begin{equation}
\label{e96}
\vL\grad\times\vE(\vr,0)=d(r,0)Y_{10}(\vOm),
\end{equation}
where the initial field amplitude $d(r,0)$ takes the explicit form
\begin{equation}
\label{e97}
d(r,0)=-2k_0^3\sqrt{4\pi\over 3}P_0{d\over dr}\int_0^{r_0} j_0(k_0 r^<) h_0^{(1)}(k_0r^>) 
r^{\prime 2}dr^\prime.
\end{equation}
The radial integral can also be performed in closed form by means of the 
indefinite-integral identities \cite{Schiff}, 
\begin{equation}
\label{e98}
\int j_0(x) x^2 dx=x^2j_1(x),\ \ \ 
\int h_0^{(1)}(x) x^2 dx=x^2h_1^{(1)}(x),
\end{equation}
and its derivative taken by means of the identities
\begin{equation}
\label{e99}
{d\over dx}j_0(x)=-j_1(x),\ \ \
{d\over dx}h_0^{(1)}(x)=-h_1^{(1)}(x).
\end{equation}
The following final form of the amplitude is obtained in this way:
\begin{equation}
\label{e100}
d(r,0)=2k_0^3r_0^2\sqrt{4\pi\over 3}P_0j_1(k_0r_0^<)h_1^{(1)}(k_0r_0^>),
\end{equation}
where $r_0^<$ and $r_0^>$ are defined to be the smaller and the larger of the two
radial distances $r$ and $r_0$, respectively.

The same steps that took us from Eq.~(\ref{e53}) to Eq.~(\ref{e55}) can be employed to write down
the displacement field envelope $\vD(\vr,t)$ in terms of the amplitude $d(r,t)$ of 
electric dipole radiation $(\ell=1)$,
\begin{equation}
\label{e101}
\vD(\vr,t)\approx {1\over 2k_0^2}\grad\times [d(r,t)\vec L Y_{10}(\vOm)].
\end{equation}
Inside the medium, the amplitude $d(r,t)$ may be expanded in terms of the mode 
functions $j_1(\gn r)\exp(-\lnn t)$, 
\begin{equation}
\label{e102}
d(r,t)= \sum_{n=1}^\infty d_n j_1(\gn r)\exp(-\lnn t).
\end{equation}
Orthogonality of the mode functions under the inner product (\ref{e90}) enables us
to solve for the coefficients, 
\begin{equation}
\label{e103}
d_p={(d(r,0),j_1(\gn r))\over 
(j_1(\gn r),j_1(\gn r))}.
\end{equation}
The inner products in Eq.~(\ref{e103}) contain integrals involving bilinear
expressions of spherical Bessel and Hankel functions of the same order 1, which,
as we show in Appendix A, can be evaluated in terms of simple trigonometric functions. 
The inner product in the numerator assumes the form,
$$(d(r,0),j_1(\gn r))=-4ik_0^2r_0^2\sqrt{4\pi\over 3}P_0{\gn^2\over k_0^2-\gn^2}
j_1(\gn r_0),$$
while that in the denominator greatly simplifies in the limit of large $|\gn|R$,
$$(j_1(\gn r),j_1(\gn r))\approx -R,$$
so the amplitude $d_n$ may be expressed as 
\begin{equation}
\label{e104}
d_n\approx 4i{k_0^2r_0^2\over R}\sqrt{4\pi\over 3}P_0{\gn^2\over k_0^2-\gn^2}
j_1(\gn r_0). 
\end{equation}

The electric field envelope $\vE(\vr,t)$ has a similar form as the electric displacement
field $\vD(\vr,t)$,
\begin{equation}
\label{e105}
\vE(\vr,t)\approx {1\over 2k_0^2}\grad\times [e(r,t)\vec L Y_{10}(\vOm)],
\end{equation}
with the amplitude $e(r,t)$ having a similar mode decomposition as $d(r,t)$,
\begin{equation}
\label{e106}
e(r,t)= \sum_{n=1}^\infty n_p j_1(\gn r)\exp(-\lnn t).
\end{equation}
The mode coefficient $e_n$ for the electric field is obtained from $d_n$, the analogous
coefficient for the displacement field, by dividing the latter by the dielectric constant 
$\gn^2/k_0^2$ for the $n$th mode, 
\begin{equation}
\label{e107}
e_n={d_n\over(\gn/k_0)^2}\approx 4i{k_0^2r_0^2\over R}\sqrt{4\pi\over 3}P_0{k_0^2\over 
k_0^2-\gn^2} j_1(\gn r_0). 
\end{equation}
The resonant character of the excitation of the modes is unmistakable in the
denominator of the right hand side of Eq.~(\ref{e107}) -- only 
modes with propagation constants $\gn$ that are closely matched to the free space  
wave vector $k_0$ are preferentially excited. Also, because $j_1(\gn r_0)$ falls off
rapidly with $\gn$ for $|\gn|r_0>>1$, only modes with propagation constants of 
order $1/r_0$ or smaller in magnitude are significantly excited in the preparation 
of the initial core excitation.

\subsection{\label{A}Frequency Spectrum of Radiation}

The initial polarization (\ref{e91}) radiates light into the various electric-field eigenmodes 
with amplitudes $e_n$. Apart from overall angular and distance dependences, the spectral 
amplitude of radiation, $a(R,\dom)$, at detuning $\dom$ is given by 
the Fourier transform of the electric field amplitude $e(R,t)$ at the surface 
of the sphere,
\begin{equation}
\label{e108}
a(\dom)=\int_0^\infty e(R,t)e^{i\dom t} dt=\sum_{n=1}^\infty {e_nj_1(\gn R)
\over \lnn-i\dom}.
\end{equation}
By substituting for $e_n$ from Eq.~(\ref{e107}) and using the relation (\ref{e33}) between $\gl$
and $\lambda$ in Eq.~(\ref{e107}), we may express the spectral amplitude as
\begin{equation}
\label{e109}
a(\dom)= 4{k_0^2r_0^2\over R}\sqrt{4\pi\over 3}P_0{\beta^2\over \dom}\sum_n
{j_1(\gn r_0)j_1(\gn R)\over \gn^2 R^2-\beta^2_\dom}, 
\end{equation}
where $\bdo$ is $R$ times the propagation constant for a plane wave of frequency
$(\omega_0+\dom)$ traversing the medium,
\begin{equation}
\label{e110}
\bdo=\beta\sqrt{1-{n_0\mus/(3\hbar)\over \dom}}.
\end{equation}
The power spectrum of radiation is obtained from $a(\dom)$ by taking its squared modulus.

The mode sum of Eq.~(\ref{e109}) may be evaluated numerically. However, because of
its resonant character which implies significant contribution from
only those modes that have propagation constant close to the plane-wave propagation
constant $\bdo$, accurate analytical approximations are also possible. First of all,
for large spheres, $\beta>>1$, since all of the mode propagation constants $\gn$ are 
largely real with relatively
small imaginary parts, there is little radiation at frequency detunings that lie
in the range $(0,n_0\mus/(3\hbar))$ for which $\bdo$ is purely imaginary. Due to
a strong resonant interaction of light with matter, radiation simply
cannot propagate far in this frequency range. There is thus
a gap in the spectrum in this range, a feature of resonant radiation that was already
noted in the one-dimensional problem \cite{PG00}. 

Outside this gap, the propagation constant $\bdo$ 
is either large or small compared to the radius parameter $\beta$, and the following
excellent analytical approximations can be verified for the propagation constants 
$\gn$ of the significantly contributing modes, those for which $|\gn|R\approx\bdo$:
\begin{align}
\label{e111}
x_n=&\gn R\nonumber\\
\approx&\begin{cases}
\left(n-{1\over 2}\right)\pi-i{\displaystyle{\beta\over \left(n-{1\over 2}\right)\pi}} 
&{\rm for\ } \dom<0 \\
n\pi-i{\displaystyle{n\pi\over\beta}} &{\rm for\ } \dom>n_0\mus/(3\hbar).
\end{cases}
\end{align}
When these expressions are substituted in the resonant denominators of the terms in the
mode sum (\ref{e109}) and in the large-argument approximation for $j_1(\gn R)$, namely
$-\cos(\gn R)/(\gn R)$, and $\gn R$ is replaced by $\bdo$ in terms that
vary slowly from one mode to the next, the following analytical forms result for
the spectral amplitude:   
\begin{widetext}
\begin{align}
\label{e112}
a(\dom)\approx &-4{k_0^2r_0^2\over R}\sqrt{4\pi\over 3}P_0{\beta^2 j_1(\bdo f)\over \dom\bdo}
\nonumber\\
&\times \begin{cases}
-i\beta{\displaystyle \sum_{n=1}^\infty{(-1)^n \over
(n-1/2)\pi\, [(n-1/2)^2\pi^2-\gdo^2]}} &{\rm for\ } \dom<0\\ 
{\displaystyle\sum_{n=1}^\infty{(-1)^n \over
n^2\pi^2-\gdo^2}} &{\rm for\ } \dom>n_0\mus/(3\hbar),
\end{cases}
\end{align}
\end{widetext}
where $f=r_0/R$ denotes the fraction of the spherical radius initially
excited and $\gdo$ an effective propagation constant that takes slightly
different forms below and above the gap, 
\begin{equation}
\label{e113}
\gdo=\begin{cases}
\sqrt{\bdo^2+2i\beta} &{\rm for\ } \dom<0\\
\bdo(1+i/\beta) &{\rm for\ } \dom>n_0\mus/(3\hbar).
\end{cases}
\end{equation}
An overall factor that differs from 1 by terms of order $1/\beta$ has been dropped
from Eq.~(\ref{e112}) which is bound to be accurate for large values of $\beta$ only. 
Both sums in Eq.~(\ref{e112}) may be evaluated in closed form.
The second sum is well known \cite{GR65}. The evaluation of the first sum by
means of contour integration is presented in Appendix B.
The following closed-form expressions are then obtained for the spectral amplitude below and 
above the frequency gap:
\begin{align}
\label{e114}
a(\dom)&\approx 4{k_0^2r_0^2\over R}\sqrt{4\pi\over 3}P_0{\beta^2 j_1(\bdo f)\over \dom\bdo}
\nonumber\\
&\times\begin{cases}
-{\displaystyle{i\beta\over 2\gdo^2}\left(1-{1\over\cos\gdo}\right)} 
&{\rm for\ } \dom<0\\ 
{\displaystyle{1\over 2\gdo}\left({1\over \sin\gdo}-{1\over \gdo}\right)} 
&{\rm for\ } \dom>{n_0\mus\over 3\hbar}.
\end{cases}
\end{align}
The squared modulus $S(\dom)=|a(\dom)|^2$ represents the power spectral density of radiation.

We plot in Figs. 4 the power spectrum $S(\dom)$ for a variety of values of $\beta$ and 
the fractional excited radius $f\equiv r_0/R$. For small spherical samples with linear dimensions 
comparable to the wavelength of light, as for $\beta=1$, coherent radiation initially
proceeds by the fast decaying superradiant mode we discussed in Sec.~II, but the
relatively slowly decaying modes that we discussed in Sec.~V.B.1 continue to radiate 
for long periods of time. Because of the simple proportionality of the decay rate
to the width of the spectrum radiated by a mode, the superradiant mode furnishes a
broad spectral background on which are coherently superposed sharper line spectra
corresponding to the slowly decaying modes. Each spectral peak corresponding to a mode is
centered at a frequency detuning equal to the imaginary part of the decay constant of
that mode. We display the power spectrum of emitted radiation for four different values of the
fractional excited core radius, $f$, namely 0.1, 0.5, 0.75, and 1, in Figs. 4(a)-(d). 
Since the more localized initial excitations are made up of a larger number of weakly 
decaying modes, all superposed coherently with the fundamental supperradiant mode, 
the overall power spectrum consists of a broad peak and a fine structure of ever narrower peaks that 
accumulate below zero detuning, as seen most dramatically in Fig.~4(d). An initial excitation of the full sphere
corresponds, by contrast, to only a small admixture of the weakly decaying modes, that are
visible in Fig.~4(a) as small peaks on the broad background provided by the superradiant mode.
When considered in sequence, these four subplots also illustrate how even for small values of
$\beta$ a frequency gap develops
over the interval $[0,n_0\mu^2/(3\hbar)]$ as the initial excitation of the sphere is more and more tightly 
confined close to its center. The broad background contribution of the superradiant mode gets progressively
smaller, yielding little power at any positive frequency detuning including this gap. 
\begin{widetext}
\begin{figure}[ht]
     \centering
     \subfigure[]
     {\label{fig:spectrum1_1}
     \includegraphics[width=.4\textwidth, height=0.18\textheight]{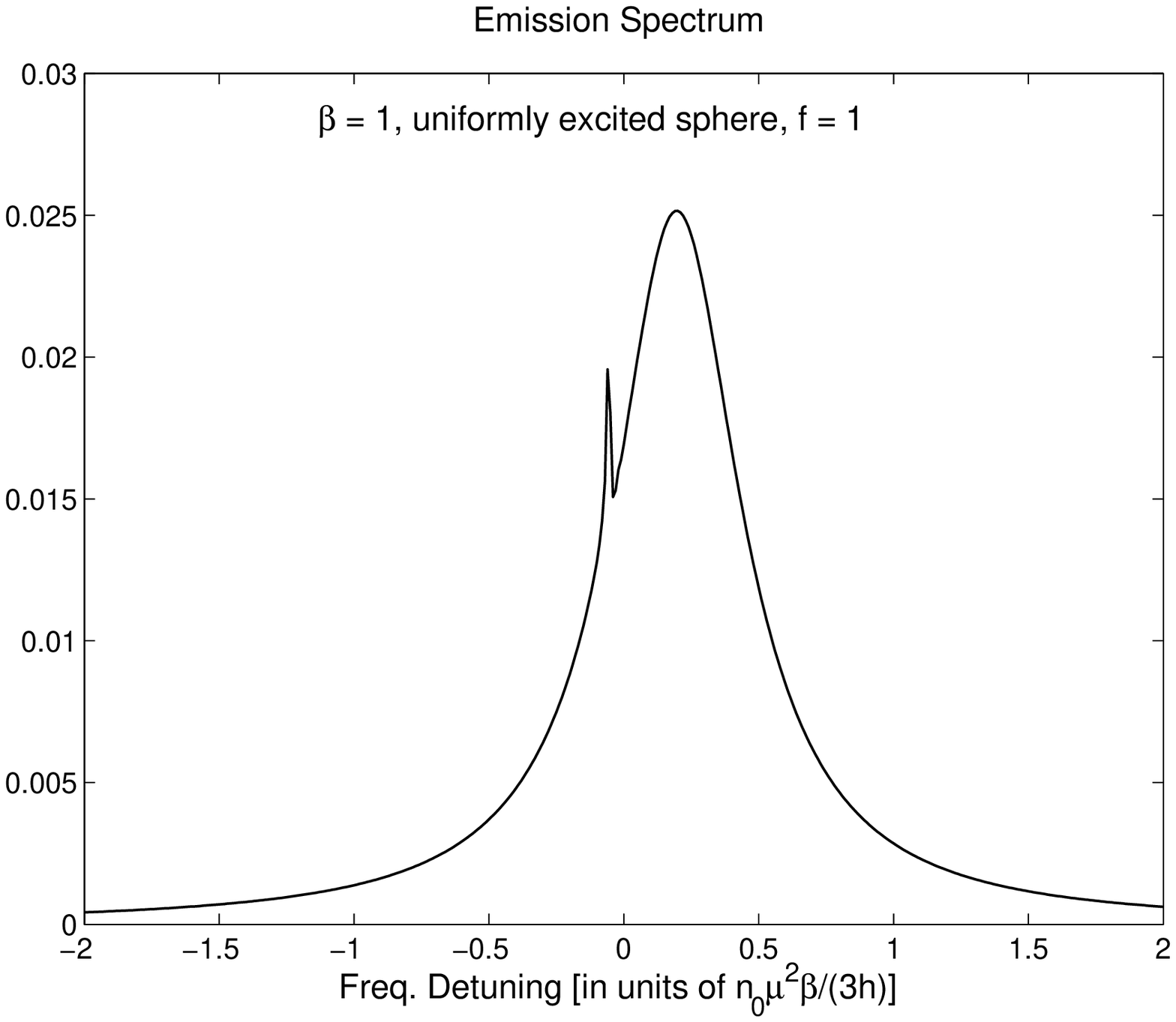}}

     \subfigure[]
     {\label{fig:spectrum1_pt75}
     \includegraphics[width=.4\textwidth, height=0.18\textheight]{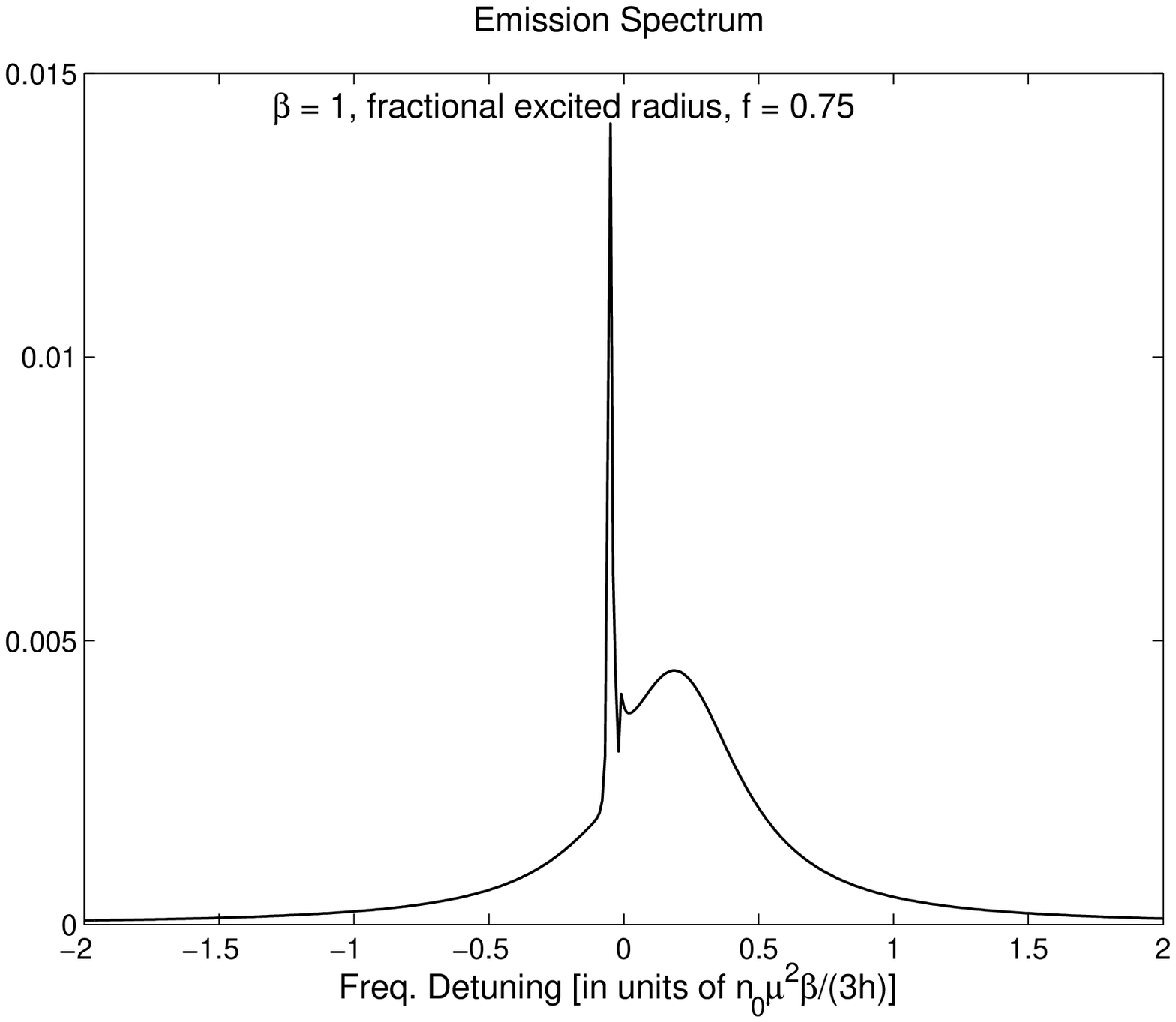}}

     \subfigure[]
     {\label{fig:spectrum1_pt5}
     \includegraphics[width=.4\textwidth, height=0.18\textheight]{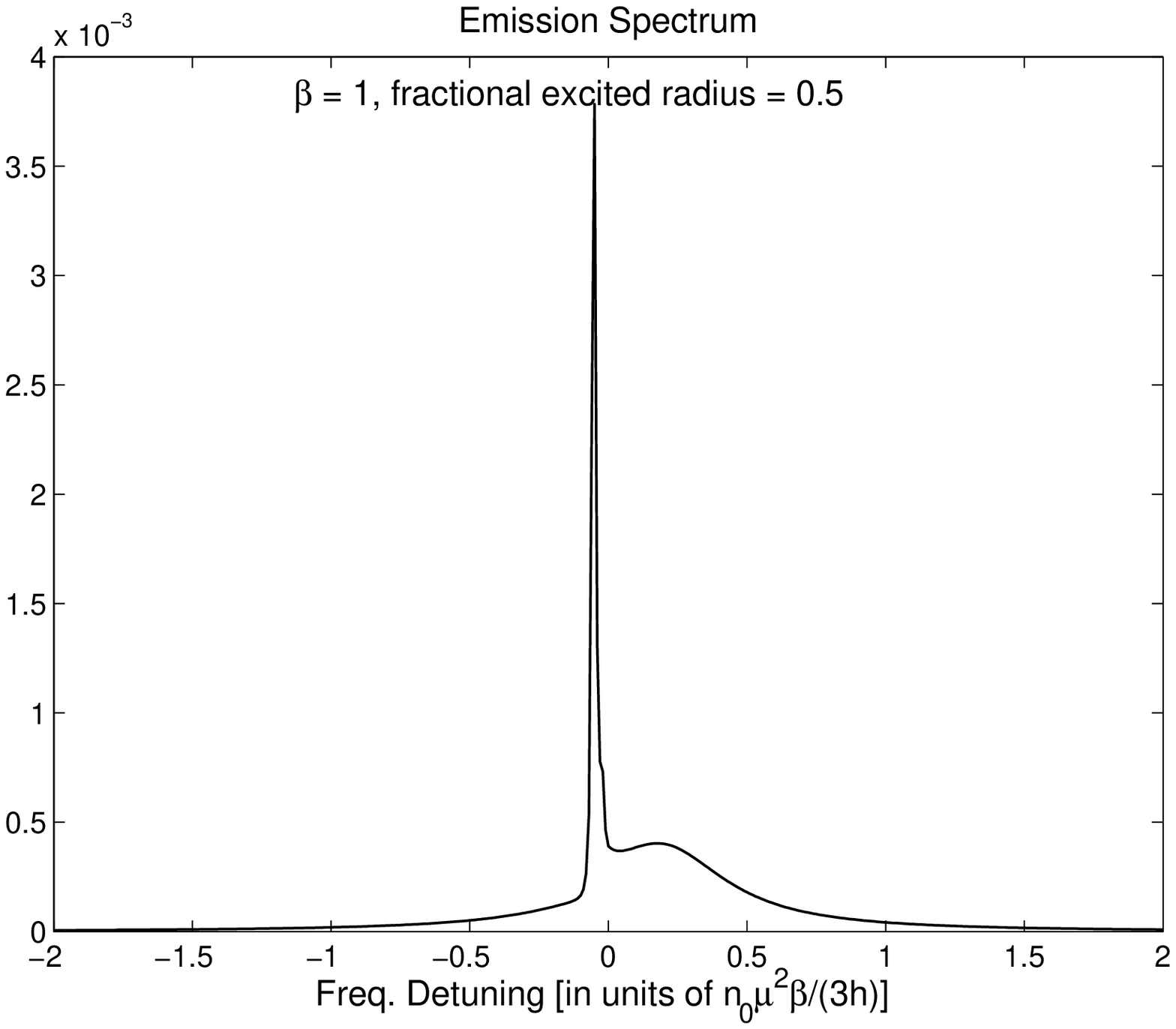}}

     \subfigure[]
     {\label{fig:spectrum1_pt1}
     \includegraphics[width=.4\textwidth, height=0.18\textheight]{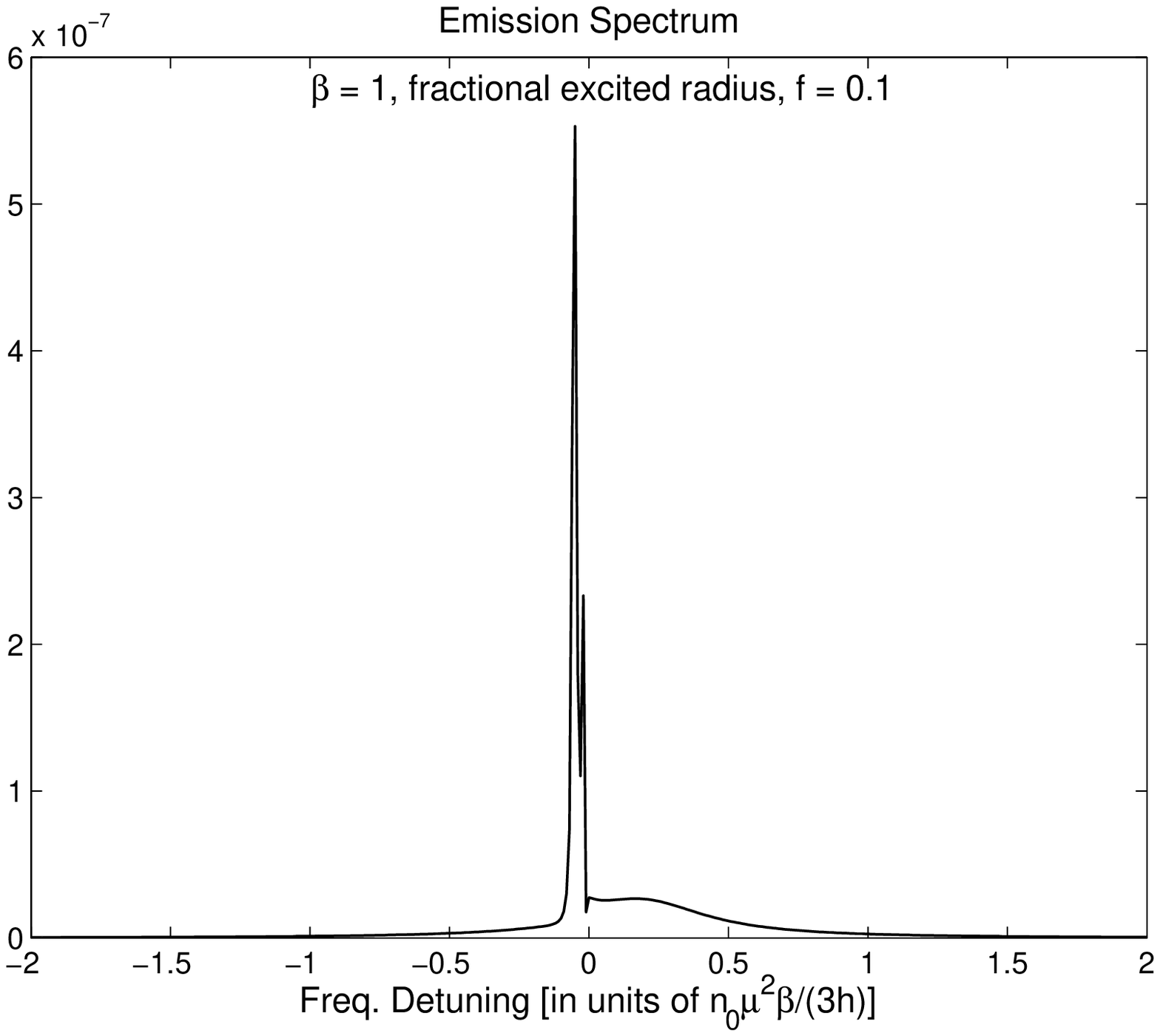}}
\caption{The frequency spectrum of the power radiated by the medium,
in arbitrary units, as a function of the frequency detuning, in units of 
$n_0\mus\beta/(3\hbar)$, for $\beta=1$, with 
(a) $f=1$ (uniform initial excitation); (b) $f=0.75$; (c) $f=0.5$; and (d) $f=0.1$} 

\end{figure}
\end{widetext}

For larger spheres too, a similar fine structure is obtained in the spectrum, but
the emergence of a well defined frequency gap we discussed earlier is unmistakable with $\beta$ 
increasing in value from 10 to 100, as we see in Figs.~5 and 6. The more localized the
initial excitation the more prominent the peaks to the left of the gap. This is a result
of the fact that the peaks with negative detuning correspond to modes that have
propagation constants that are large compared to $k_0$ and which are therefore needed to
make up an excitation that is localized on the sub-wavelength scale corresponding to
$\beta f<<1$. The analytical approximation (\ref{e114}), shown by a dashed curve
on each plot, is already accurate for $\beta=10$, and is nearly indistinguishable from
the numerically exact results for $\beta=100$.
For $\beta=10$ the most superradiant mode has a decay rate that, according to expression (\ref{e76}),
is a factor $\ln \beta =2.3$ times smaller than $n_0\mu^2\beta/(3\hbar)$. 
This accounts for the considerably narrower background provided
by this mode to the emission spectrum in Fig.~5(a) than that present in Fig.~4(a) for $\beta=1$, 
even though that mode dominates the other modes when the sphere is uniformly excited. 
For much larger values of $\beta$, as in Figs.~6, the superradiant modes, which are
of order $(\ln\beta)/\pi$ in number, each correspond to a spatially nonuniform excitation given by Eq.~(\ref{e51})
with $\gl R\sim\beta$. A uniformly excited core, regardless of its fractional radius, $f$, 
is thus comprised of large numbers of superradiant and weakly decaying modes, 
which always yield a rich spectrum of narrow peaks that accumulate on either side of the
frequency gap. 

\begin{widetext}
\begin{figure}[ht]
     \centering
     \subfigure[]
     {\label{fig:spectrum10_1}
     \includegraphics[width=.4\textwidth, height=0.25\textheight]{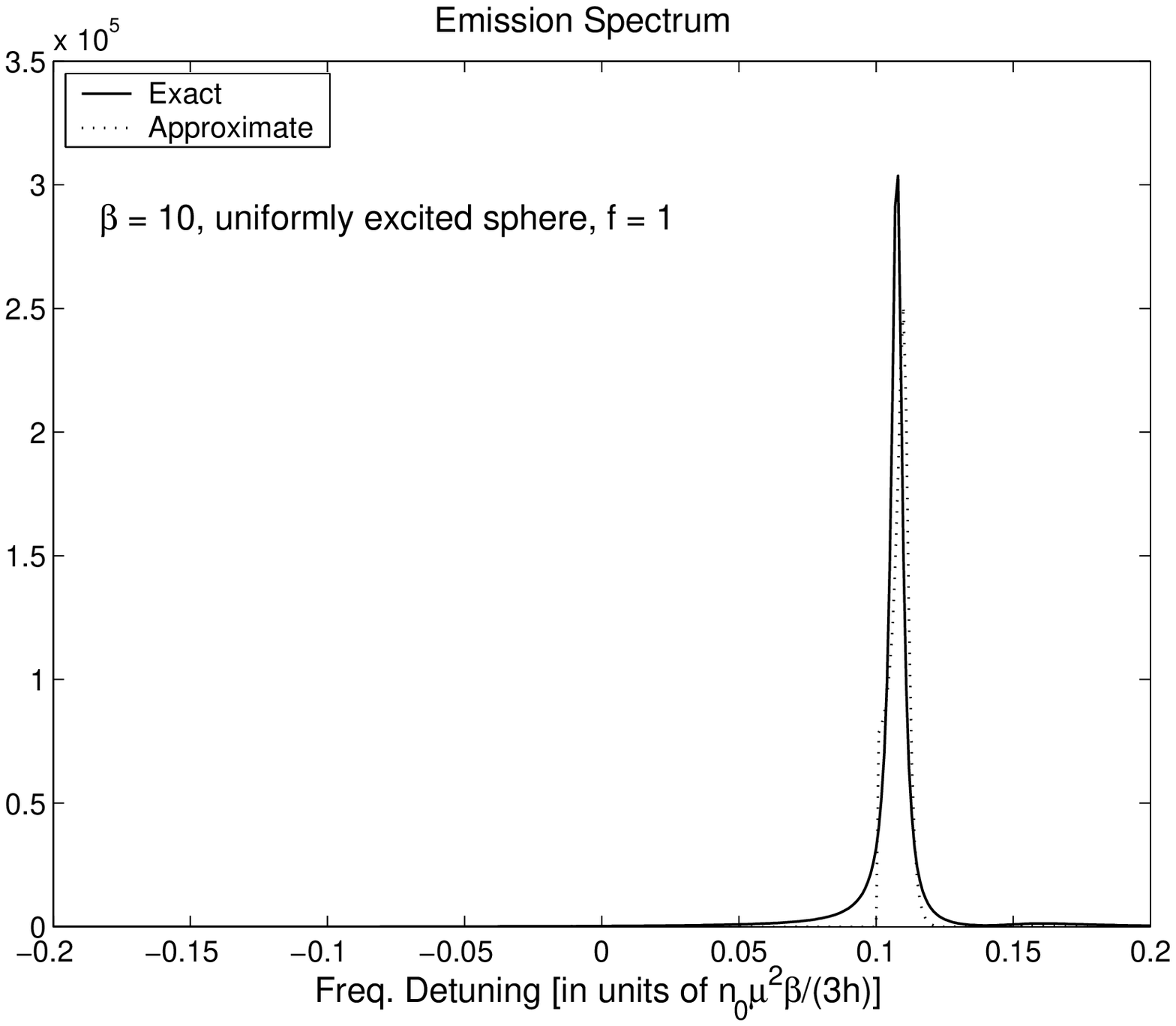}}

     \subfigure[]
     {\label{fig:spectrum10_01}
     \includegraphics[width=.4\textwidth, height=0.25\textheight]{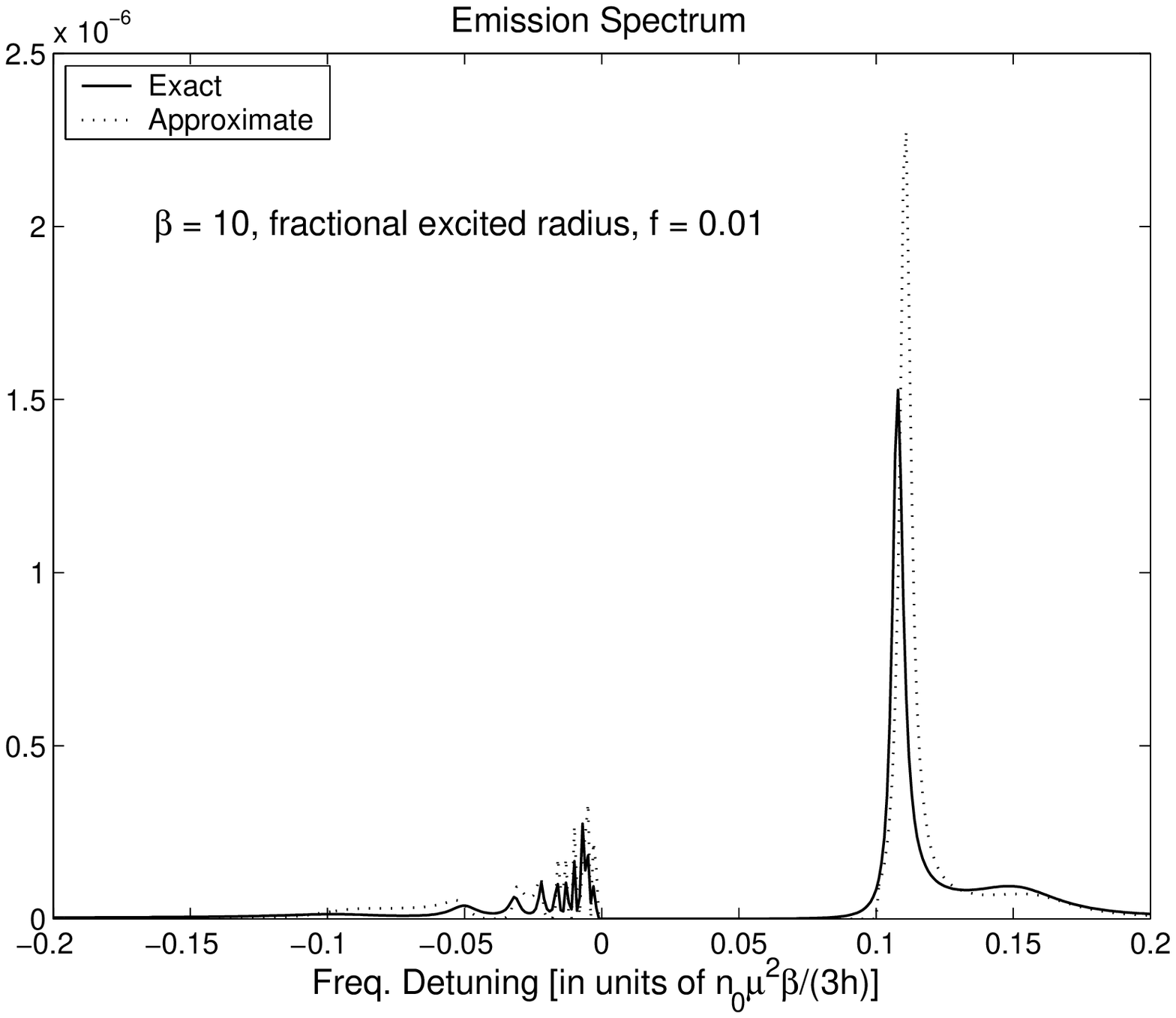}}
\caption{Same as Fig. 4(a), except (a) $\beta=10$, $f=1$ and (b) $\beta=10$, $f=0.01$.}
\end{figure}

\begin{figure}[ht]
     \centering
     \subfigure[]
     {\label{fig:spectrum100_01}
     \includegraphics[width=.4\textwidth, height=0.25\textheight]{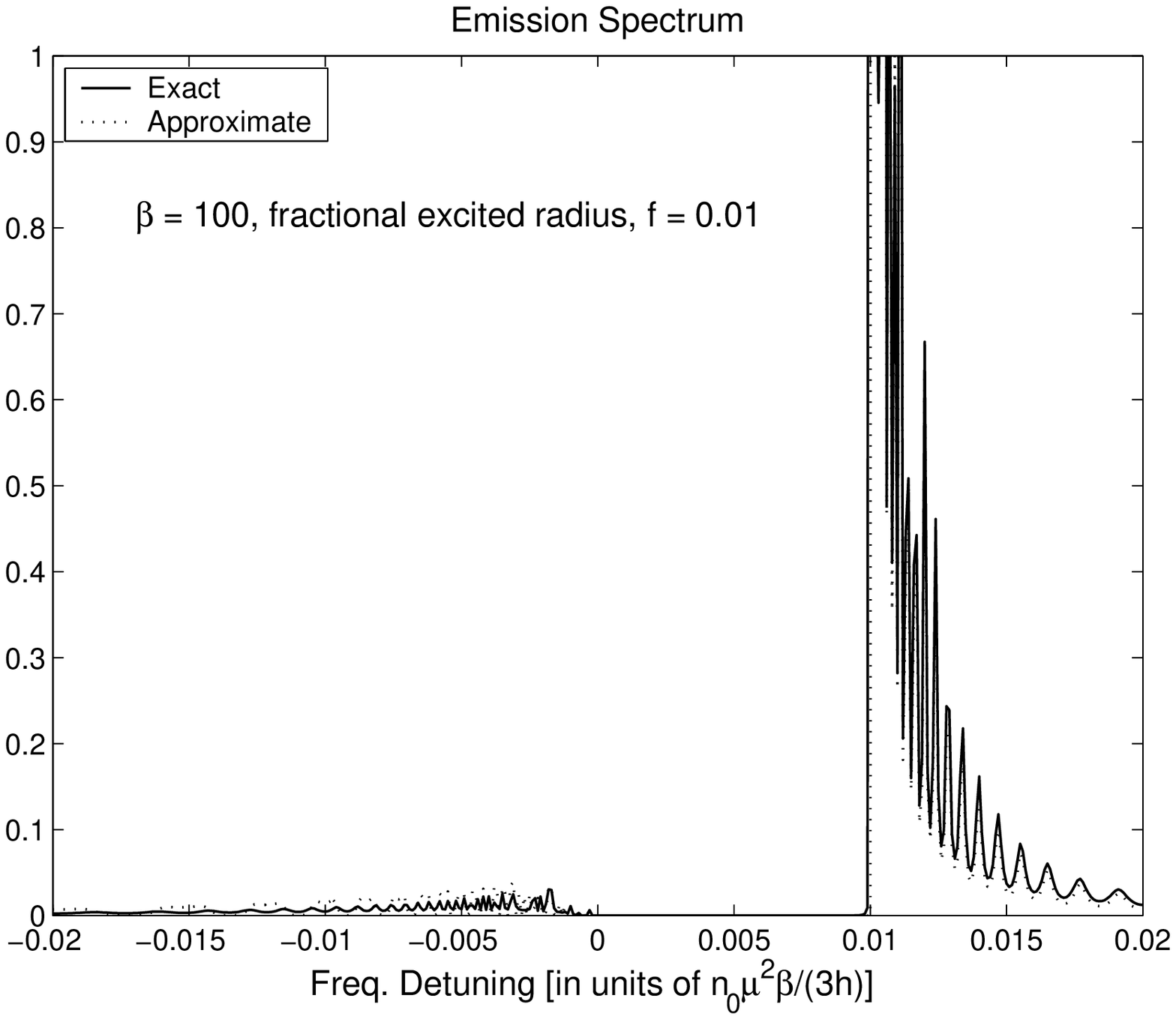}}

     \subfigure[]
     {\label{fig:spectrum100_0001}
     \includegraphics[width=.4\textwidth, height=0.25\textheight]{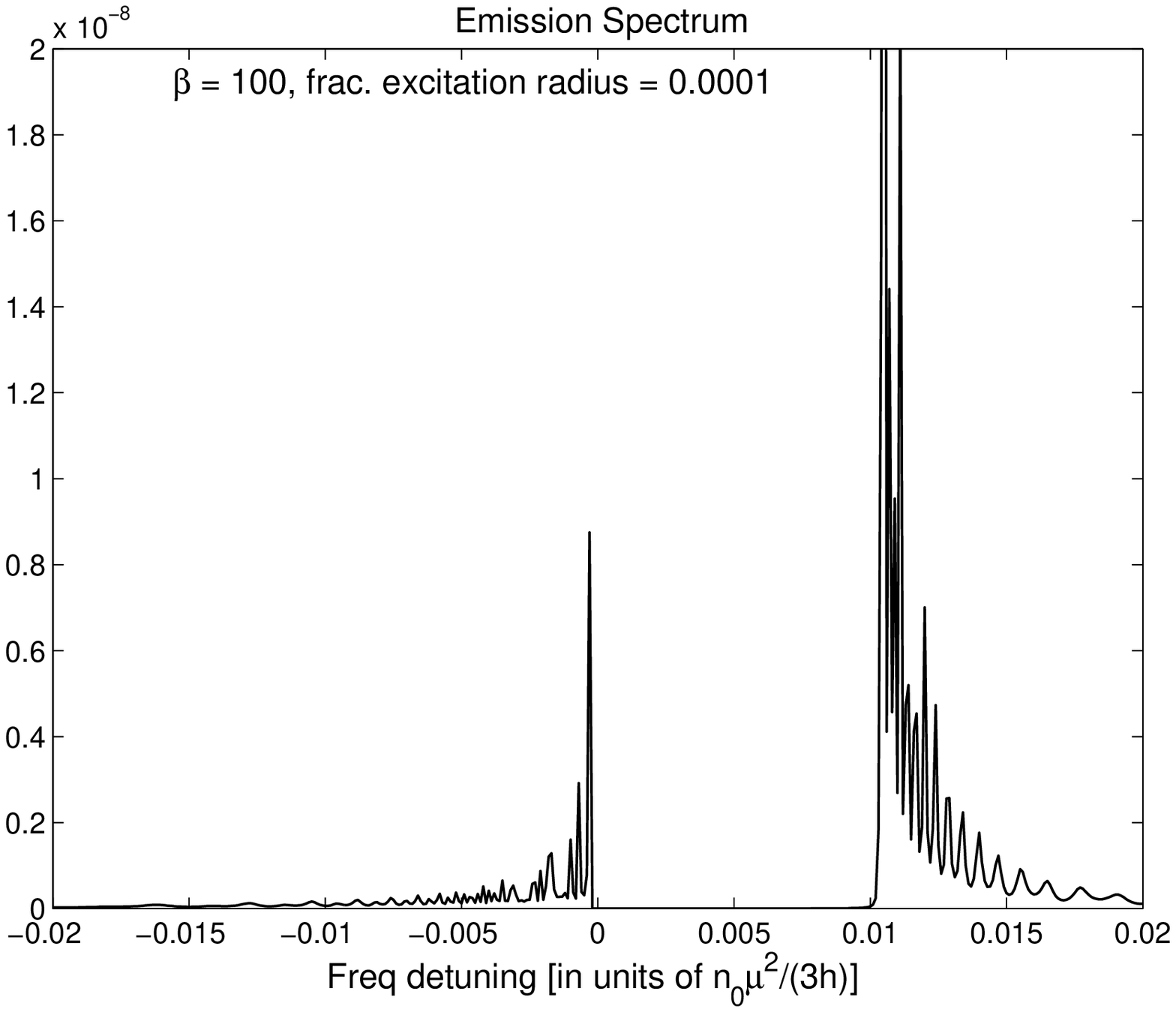}}
\caption{Same as Figs. 5(a), except (a) $\beta=100$, $f=0.01$ and (b) $\beta=100$,
$f=0.0001$.}
\end{figure}
\end{widetext}

\subsection{\label{B}Time Dependence of the Radiated Power}

The power $dW/dt$ radiated by the resonantly excited sphere is given by integrating the
normal component of the Poynting vector over the surface of the sphere.
For $\beta>>1$, this integral may be reduced to a rather simple final form, as we show
in Appendix C,
\begin{equation}
\label{e115}
{dW\over dt}=-{c\beta^4\over k_0^4(1+\beta^2)}|d(R,t)|^2,
\end{equation}
where $d(R,t)$ is the amplitude function that determines the electric displacement field
at the spherical surface, $r=R$, via relation (\ref{e102}). 

\subsubsection{Radiation from a Large Sphere, $\beta>>1$, with a Small Excited Core,
$f<<1$}

The early stages of radiation are dominated by the fast-decaying, 
superradiant modes, while the excitation residing in the more weakly radiating modes 
is slow to escape the medium. The latter are accompanied by an oscillatory
exchange of energy between the atoms and the radiation field, as we
shall see presently. Similar attributes of cooperative emission 
from extended media have also been seen for a symmetric
Dicke one-photon excitation \cite{SS091,M09}.  

If we use in expression (\ref{e104}) the approximate eigenvalue 
formulas (\ref{e75}) and (\ref{e76}), valid for the superradiant modes of a large sphere, 
and then evaluate the sum (\ref{e102}), we are assured of good accuracy for the times when 
those modes are the ones actively radiating.
Since the superradiant modes have propagation constants that do not differ much from
the free-space one, $k_0$, we first approximate $(k_0^2-\gn^2)$ by 2$k_0(k_0-\gn)$, 
$\gn^2$ by $k_0^2$, 
and then substitute formulas (\ref{e75}) and (\ref{e76}) in the expression (\ref{e104}). We assume here
a small initially excited core, $r_0<<R$, so we may replace $j_1(\gn r_0)$ by 
$j_1(\beta f)$, while the large-argument approximation for $j_1(\gn R)$ coupled
with expression (\ref{e75}) reduces it to the form
\begin{equation}
\label{e116}
j_1(\gn R)\approx -{\cos \gn R\over \gn R}\approx (-1)^{n+1}{e^{i\pi/4}\over 
\sqrt{\beta}},
\end{equation}
where we used $\beta>>1$ to ignore a term of relative order $1/\beta$ and replaced
$\gn R$ in the denominator simply by $\beta$.
With these approximations and upon extending the sum (\ref{e102}) to also include all negative 
integral values of the index $n$, which adds spurious 
terms to the sum that we later subtract out approximately, we may express the 
field amplitude $d(R,t)$ at the surface of the sphere as
\begin{align}
\label{e117}
d(R,t)=& -ie^{i\pi/4}\sqrt{4\pi\over 3}{f^2\beta^2}j_1(\beta f)P_0\nonumber\\
&\times\sum_{n=-\infty}^\infty {(-1)^n e^{\displaystyle {i\alpha t/[n\pi-\beta-(i/2)\ln \beta]}}
\over [n\pi-\beta-(i/2)\ln \beta]},
\end{align}
This sum may be evaluated exactly by the method of contour
integrals in the complex plane, as shown in Appendix D, and the following asymptotic
expansion in powers of $1/\beta$ is obtained:
\begin{align}
\label{e118}
d(R,t)=& 2e^{i\pi/4}\sqrt{4\pi\over 3}{f^2\beta^2}j_1(\beta f)P_0
{e^{i\beta}\over\sqrt{\beta}}\nonumber\\
&\times\sum_{n=0}^\infty \left({e^{2i\beta}\over \beta}\right)^n
J_0(2\sqrt{(2n+1)\alpha t}),
\end{align}
where $J_0$ denotes Bessel function of the first kind of order 0.
For large values of $\beta$, the first term alone suffices to furnish an accurate 
result for $d(R,t)$ and thus for the radiated power (111),
\begin{equation}
\label{e119}
{dW\over dt}\approx -{16\pi\over 3}\omega_0 k_0r_0^4 [j_1(\beta f)]^2|P_0|^2
J_0^2(2\sqrt{\alpha t}).
\end{equation}
The oscillatory time dependence (\ref{e119}) of the radiated power represents an exchange
of energy between the field and the polarization of the radiating medium. Such
oscillatory energy exchange is characteristic of any radiation problem
in which many modes of comparable decay rates that are detuned by different amounts from 
the resonance frequency participate coherently. A squared Bessel
function time dependence related to Eq.~(\ref{e119}) was first derived by Burnham and Chiao
\cite{BC69} in the one-dimensional context of radiation from a semi-infinite medium that is
coherently generated in the wake of a sweeping $\delta-$function excitation pulse.

An expression for the radiated power that is somewhat more accurate than Eq.~(\ref{e119})
is given by subtracting out the spurious terms, those with $n$ running from
0 to $-\infty$, that we included in the sum (\ref{e117}) in order to derive the Bessel-function
result (\ref{e118}). Because of their non-resonant character, the sum of
these spurious terms can be evaluated approximately, as we show in Appendix E, and 
subtracted from Eq.~(\ref{e117}). This procedure yields the following two-term result for 
the radiated power: 
\begin{align}
\label{e120}
{dW\over dt}\approx& -{16\pi\over 3}\omega_0 k_0r_0^4 [j_1(\beta f)]^2|P_0|^2
\Bigg[J_0^2(2\sqrt{\alpha t})\nonumber\\
&- {1\over 2\sqrt{\beta}}J_0(2\sqrt{\alpha t})
\sin(\beta+\alpha t/\beta)e^{-\displaystyle{{\alpha t\over 2\beta^2}\ln\beta}}
\Bigg].
\end{align}

In Figs. 7-9, we plot the power radiated by a spherical medium for the same values of
the radius parameter $\beta$ for which the
radiated spectrum was considered in Figs. 4-6. For a uniformly excited sphere 
($f=1$) with $\beta=1$, the fundamental superradiant mode is nearly the only one 
excited, which implies a purely exponential decay of the radiated power, as seen in 
Fig.~7(a). When the medium is initially excited from the center
out to only a tenth of the radius, as in Fig.~7(b), 
a significant number of weakly decaying modes of comparable magnitude
are present as well. The initial
precipitous drop in power results from the decay of the superradiant mode, but the 
subsequent emission has an oscillatory time dependence due to radiation emitted coherently by 
the weakly decaying modes. The oscillations are slow with a long quasi-period, due to rather 
small differences in the frequency detunings of these modes, as we noted earlier. 

\begin{widetext}
\begin{figure}[ht]
     \centering
     \subfigure[]
     {\label{fig:time_dep1_1}
     \includegraphics[width=.4\textwidth, height=0.25\textheight]{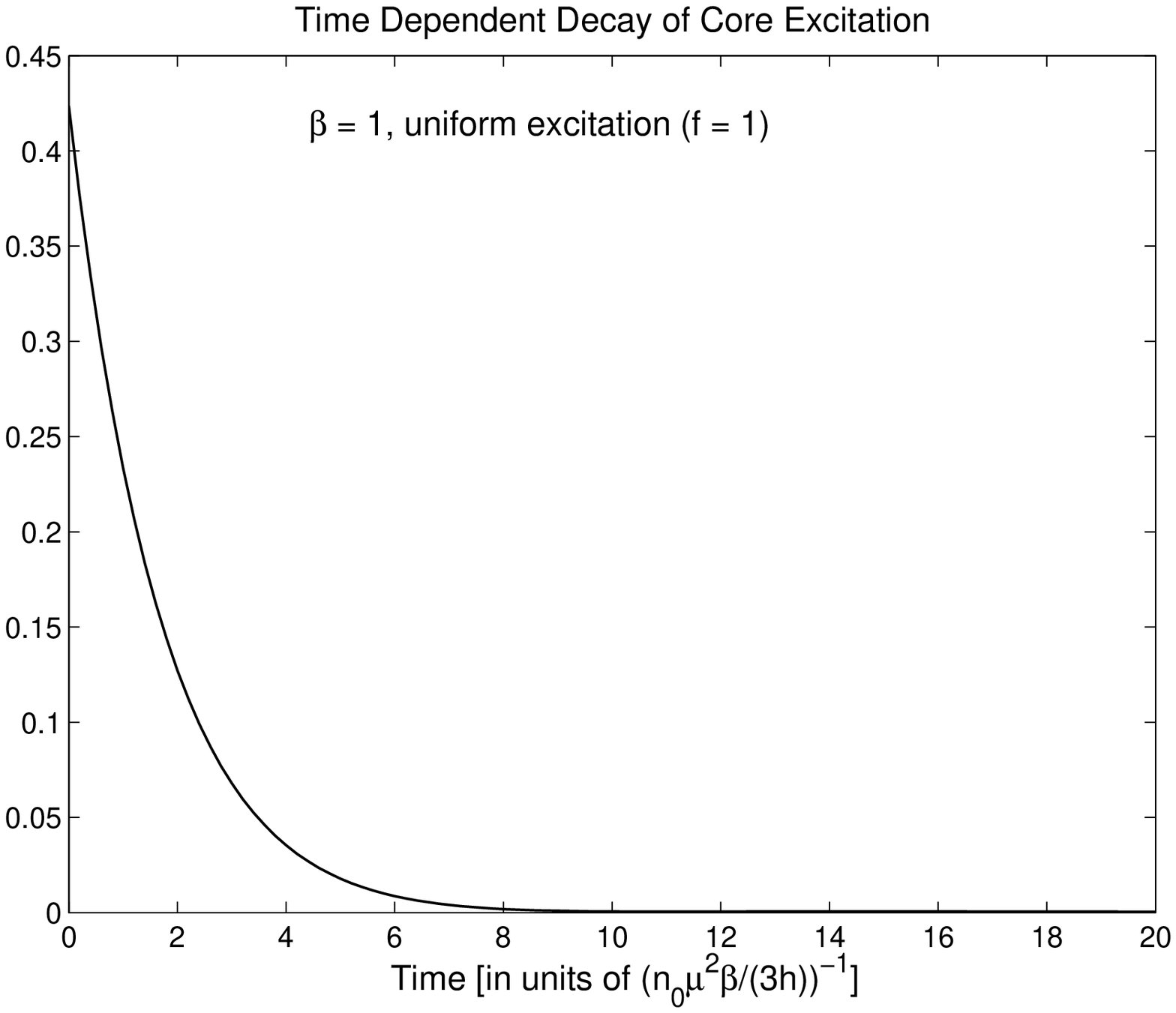}}

     \subfigure[]
     {\label{fig:spectrum1_pt1}
     \includegraphics[width=.4\textwidth, height=0.25\textheight]{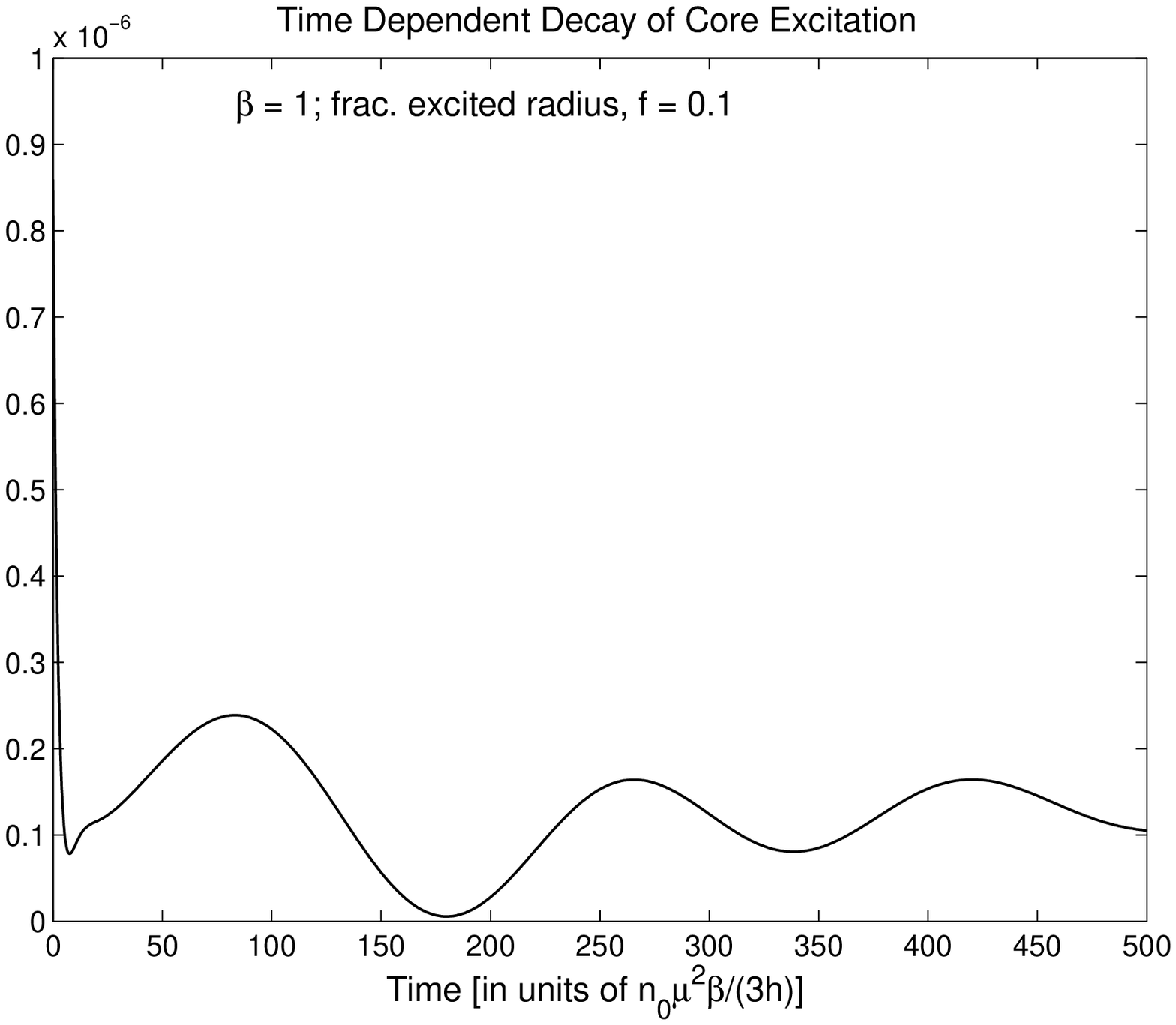}}
\caption{The power radiated by the medium, in arbitrary units, as a function of time, 
in units of $[n_0\mus\beta/(3\hbar)]^{-1}$, for $\beta=1$ with (a) $f=1.0$ and (b) 
$f=0.1$.}
\end{figure}
\end{widetext}

As we increase the value of $\beta$, we find inevitably the emergence of
an oscillatory time dependence for power that for early times is described approximately by the Bessel 
function result (\ref{e120}). This is particularly accurate for the larger
value, 100, of $\beta$ when the fractional excited core radius, $f$, is small, as seen 
in Figs. 9(b). The approximate result (\ref{e120}) ceases, however, to agree with the 
numerically exact behavior at long times due 
to the fact that the former wrongly presupposes that the superradiant modes continue to 
dominate the radiation for all times. In reality, the weakly decaying modes continue
to radiate for long times, well after the superradiant modes have radiated away nearly
all of their excitation. The differences between the frequencies of neighboring 
modes, that determine the detailed character of the temporal oscillations, are dissimilar
for the two varieties of modes, accounting for the failure of the approximate result 
(\ref{e120}) to describe the long-time behavior of the radiated power. It is worth noting that
times greatly exceeding those plotted would be necessary to see the stretched exponential
decay of polarization energy predicted in the one-dimensional problem of localized
excitation \cite{PG00}.
\begin{widetext}
\begin{figure}[ht]
     \centering
     \subfigure[]
     {\label{fig:time_dep10_1}
     \includegraphics[width=.4\textwidth, height=0.25\textheight]{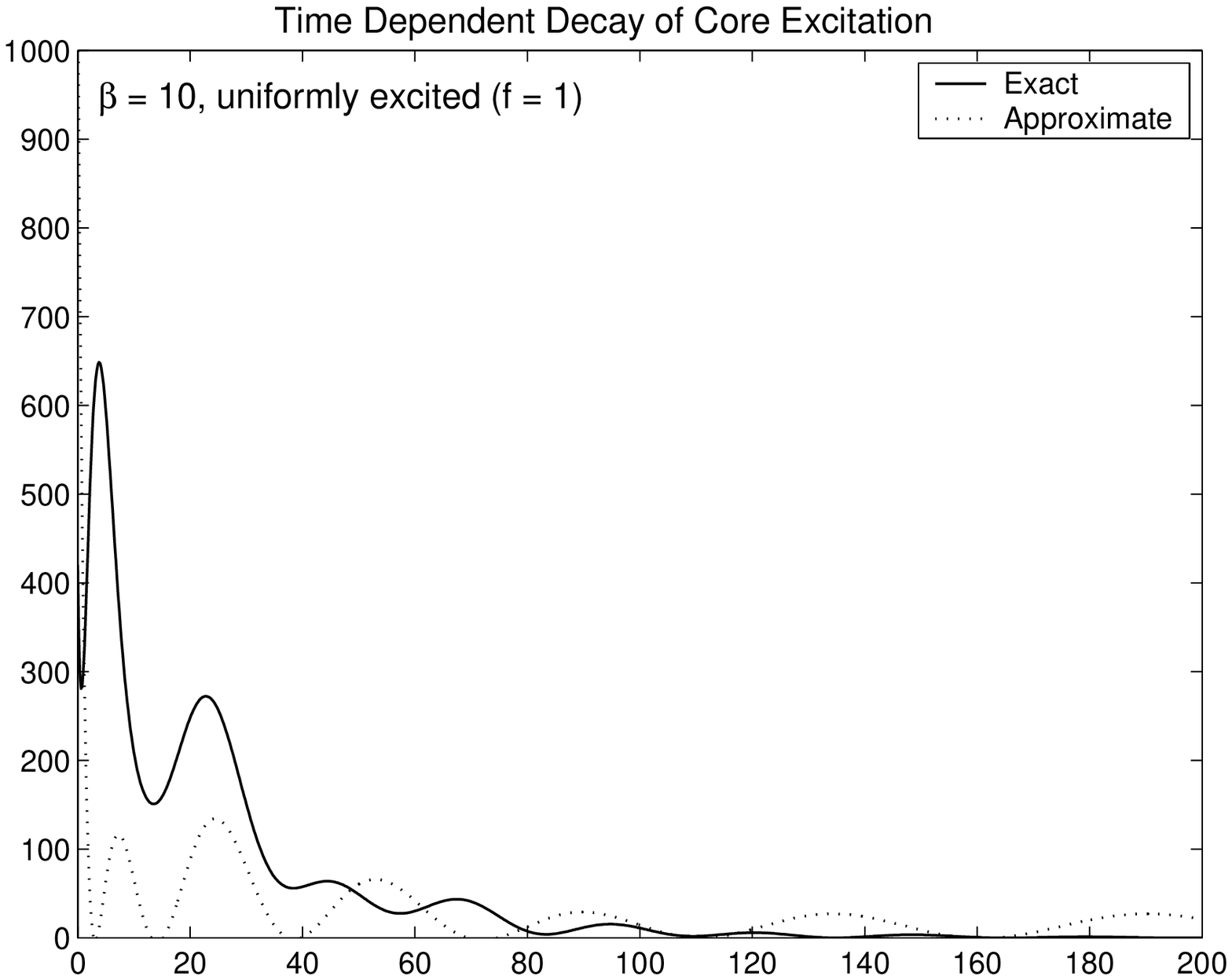}}

     \subfigure[]
     {\label{fig:spectrum1_pt1}
     \includegraphics[width=.4\textwidth, height=0.25\textheight]{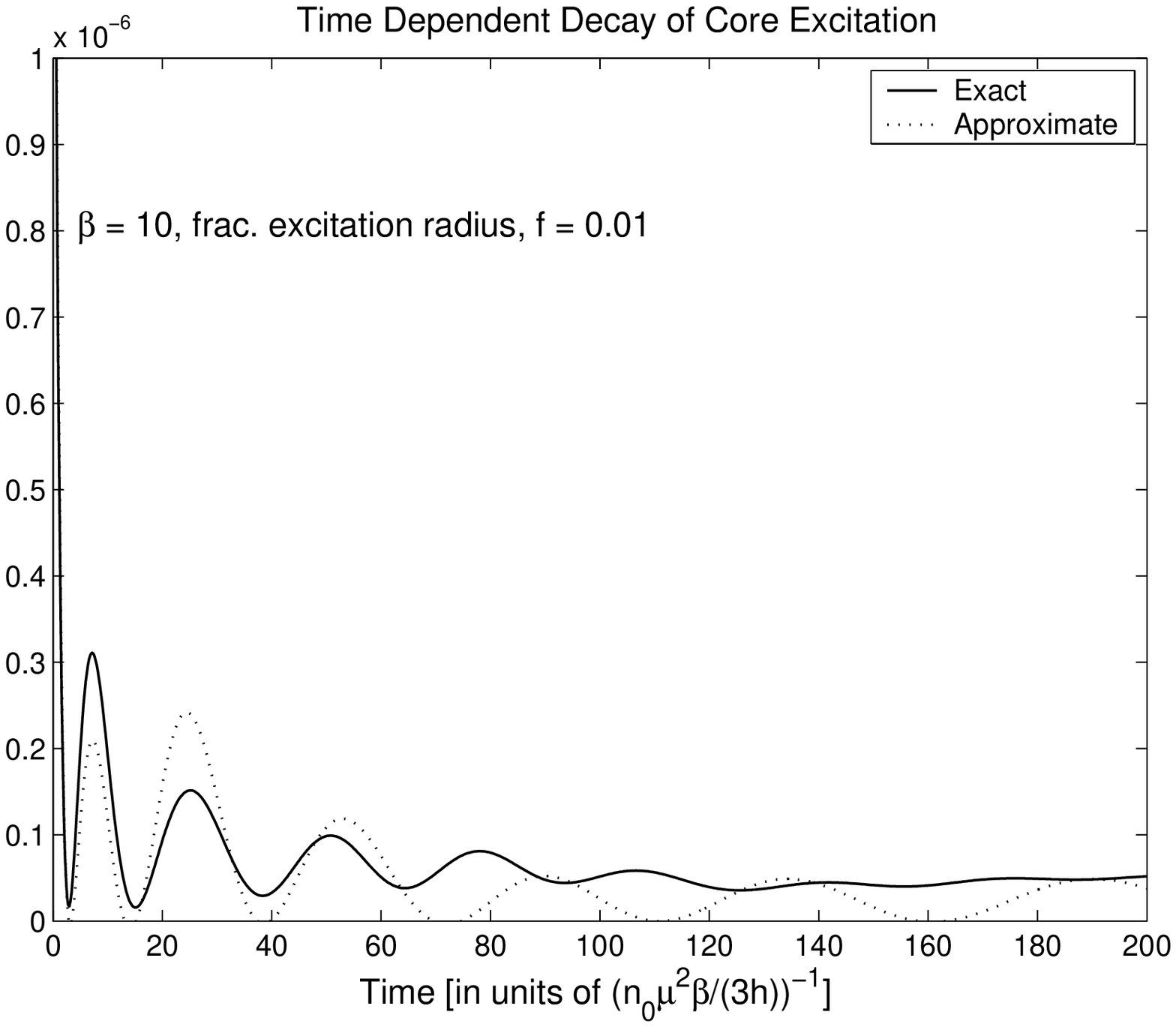}}
\caption{Same as Figs. 7(a) and (b), except (a) $\beta=10$, $f=1.0$ and (b) $\beta=10$,
$f=0.01$.}
\end{figure}

\begin{figure}[ht]
     \centering
     \subfigure[]
     {\label{fig:time_dep10_1}
     \includegraphics[width=.4\textwidth, height=0.25\textheight]{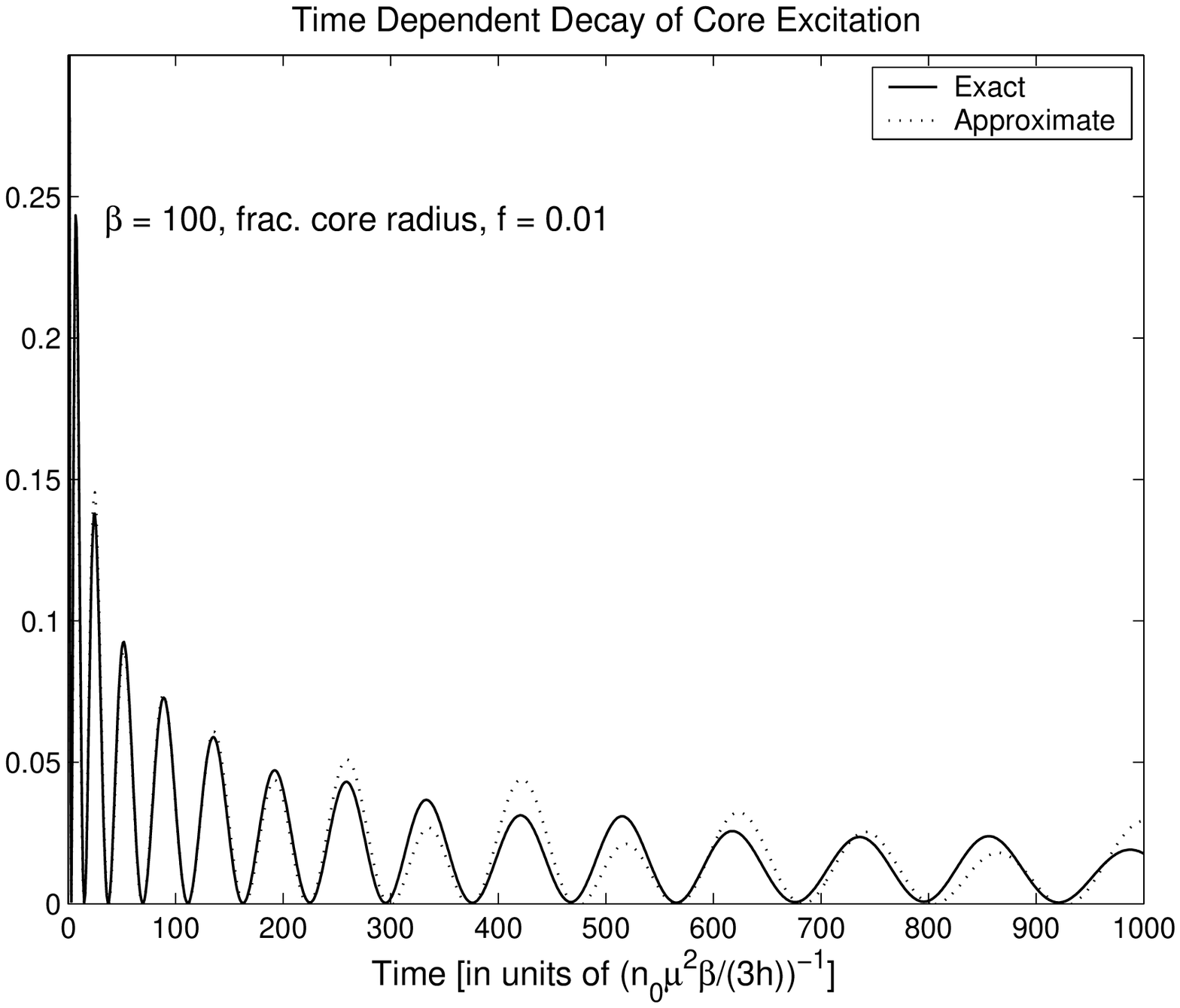}}

     \subfigure[]
     {\label{fig:spectrum1_pt1}
     \includegraphics[width=.4\textwidth, height=0.25\textheight]{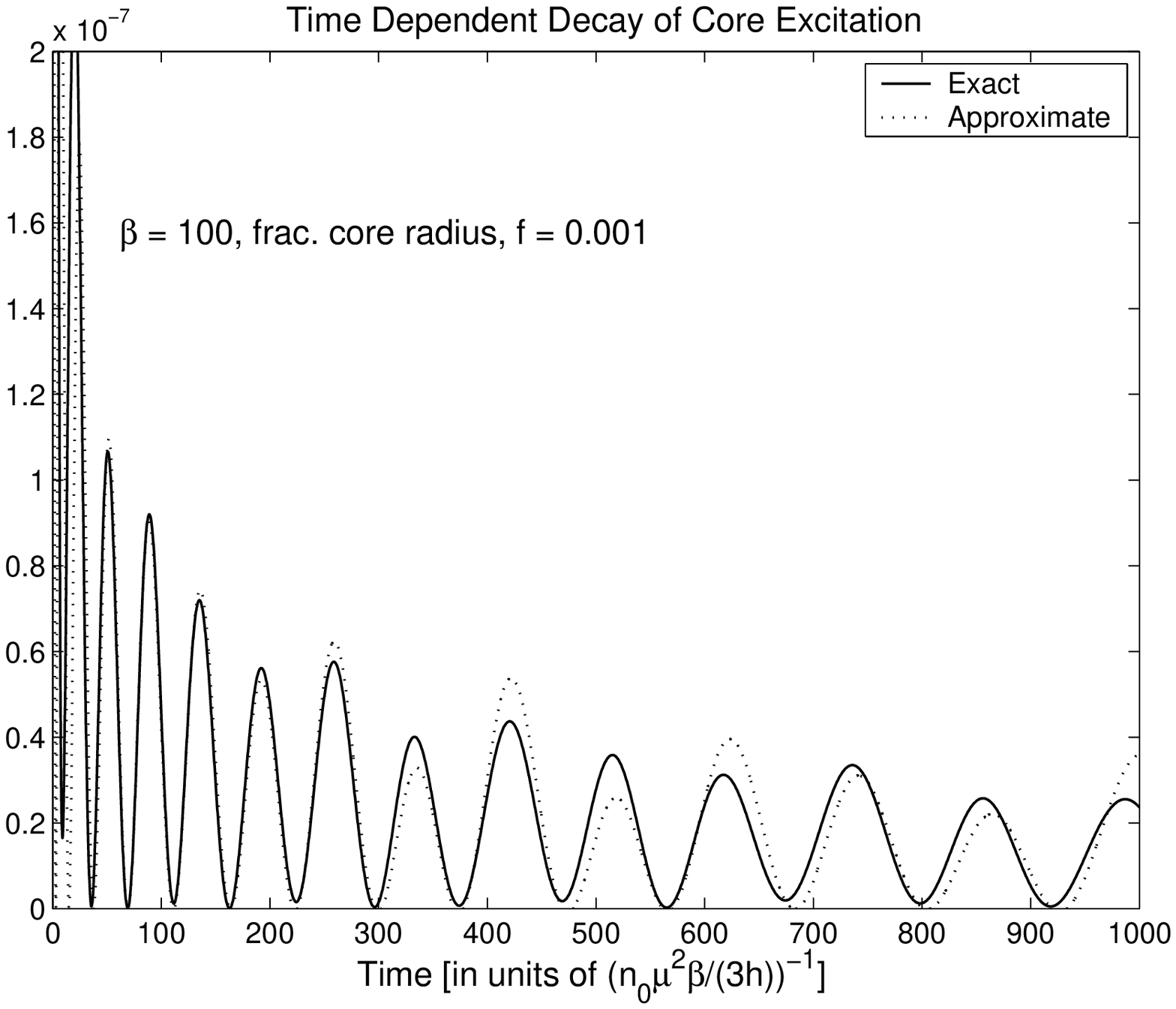}}
\caption{Same as Figs. 7(a) and (b), except (a) $\beta=100$, $f=0.01$ and (b) $\beta=100$,
$f=0.001$.}
\end{figure}
\end{widetext}

\section{\label{VIII}Concluding Remarks}

Coherent radiation from a sphere of polarizable atoms with a single resonant excitation energy level can exhibit
a rich variety of spectral and temporal characteristics. A fully excited sphere with 
uniform polarization radiates much like a single atom, when
its radius is small compared to the wavelength of emission. When
coherently excited, a large sphere, by contrast, almost always radiates a superposition
of strongly decaying, or superradiant, and weakly decaying, or subradiant, modes. These
exponentially decaying modes may be classified as magnetic and electric
multipole modes of various angular momentum orders. A particularly simple, yet 
sufficiently general, case of our problem is that of radiation from a uniformly
excited concentric spherical core within an otherwise unexcited spherical medium.
We have demonstrated that such an excitation radiates
as a pure electric dipole, regardless of the size of the medium or its excited core.
This specific case has been studied in detail in the present paper.

The characteristic differences between the frequency detunings and decay 
rates of the modes give rise to radiation that has a wealth of 
sharp peaks and valleys in its spectrum. It also has a frequency range in which
radiation cannot propagate far because of its strong resonant interaction within the medium,
and is thus unable to escape it. We derived an analytical approximation of this 
spectrum that involves simple trigonometric functions and is
highly accurate for large values of $\beta$.
The time dependence is correspondingly quite involved, with a complicated oscillatory 
behavior that persists for long times. For times that are not too long, a simple
analytical result based on the inclusion of superradiant modes alone furnishes
a good approximation to the exact time dependence.
Many of these characteristics of emission, previously noted in our treatment of 
the one-dimensional coherent radiation problem as well, are likely to survive the change
of geometry, provided geometrical length scales remain large compared to the characteristic
wavelengths of emission. 
A well localized excitation deep in the interior of an extended 
medium of arbitrary geometry will, for example, remain
trapped for long periods of time, releasing energy only slowly, unless incoherent
processes intervene. 
 
\appendix
\section{Evaluation of the Inner Products in Eq.~(99)}

When expression (\ref{e100}) for $d(r,0)$ is substituted into the inner product formula (86),
we obtain
\begin{align}
\label{A1}
(d(r,0),&j_1(\gn r))=-4k_0^3r_0^2\sqrt{4\pi\over 3}P_0\nonumber\\
&\times\Big\{Rj_1(k_0 r_0)h^{(1)}_1(k_0R)
[Rj_1(\gn R)]^\prime\nonumber\\
&\qquad+\gn^2\int_0^R r^2 h^{(1)}_1(k_0r^>)j_1(k_0r^<)j_1(\gn r) dr\Big\},
\end{align}
where a prime superscript denotes a derivative with respect to the radial coordinate,
here $R$. Since $r^>$ is the
larger of $r,r_0$ and $r^<$ the smaller, to evaluate the integral in Eq.~(A1), we write
it as a sum of two integrals
\begin{align}
\label{A2}
h_1^{(1)}(k_0r_0)&\int_0^{r_0}r^2j_1(k_0r)j_1(\gn r) dr\nonumber\\
&+ j_1(k_0r_0)\int_{r_0}^R r^2 h^{(1)}_1(k_0r^)j_1(\gn r) dr.
\end{align}
Indefinite integrals of the form $\int z_\ell(\alpha r)w_\ell(\beta r) r^2 dr$,
where $z_\ell$ and $w_\ell$ are any two spherical Bessel functions of order $\ell$,
can be computed analytically. To see how, we first recognize that Bessel functions 
obey appropriate Bessel differential equations,
\begin{equation}
\label{A3}
[rz_\ell(\alpha r)]^{\prime\prime}+[\alpha^2-\ell(\ell+1)/r^2][rz_\ell(\alpha r)]=0,
\end{equation}
\begin{equation}
\label{A4}
[rw_\ell(\beta r)]^{\prime\prime}+[\beta^2-\ell(\ell+1)/r^2][rw_\ell(\beta r)]=0.
\end{equation}
By multiplying Eq.~(\ref{A3}) by $rw_\ell(\beta r)$ and Eq.~(\ref{A4}) by $rz_\ell(\alpha r)$, then
subtracting one resulting equation from the other, and finally integrating both sides
over $r$ followed by a simple rearrangement of terms, we obtain 
\begin{align}
\label{A5}
\int r^2 z_\ell(\alpha r)w_\ell(\beta r) r^2 dr =& {1\over (\alpha^2-\beta^2)}\nonumber\\
&\times\int dr 
[rz_\ell(rw_\ell)^{\prime\prime}-rw_\ell(rz_\ell)^{\prime\prime}].
\end{align}
Since the integrand of the right hand side of Eq.~(\ref{A5}) is the derivative of 
$[rz_\ell(rw_\ell)^\prime-rw_\ell(rz_\ell)^\prime]$, its integral is trivially evaluated,
and Eq.~(\ref{A5}) reduces to the form
\begin{equation}
\label{A6}
\int r^2 z_\ell(\alpha r)w_\ell(\beta r) r^2 dr = {1\over (\alpha^2-\beta^2)}
[rz_\ell(rw_\ell)^\prime-rw_\ell(rz_\ell)^\prime].
\end{equation}
The indefinite-integral formula (\ref{A6}) may now be used to compute the 
two definite integrals in expression (\ref{A2}), and thus the integral in Eq.~(\ref{A1}) for which
the following expression results:
\begin{widetext}
\begin{eqnarray}
\label{A7}
&\int_0^R& r^2 h^{(1)}_1(k_0r^>)j_1(k_0r^<)j_1(\gn r) dr\nonumber\\
&=&{1\over (k_0^2-\gn^2)}\Big\{j_1(k_0r_0)\left[Rh_1^{(1)}(k_0R)(Rj_1(\gn R))^\prime
-Rj_1(\gn R)(Rh_1^{(1)}(k_0R))^\prime\right] \nonumber\\
 &+& j_1(\gn r_0)\left[r_0j_1(k_0r_0)(r_0h^{(1)}_1(k_0r_0))^\prime-r_0h^{(1)}_1(k_0r_0)
(r_0j_1(k_0r_0))^\prime\right]\Big\}.
\end{eqnarray}
\end{widetext}
Use of a Wronskian identity turns the terms within the second pair of brackets in Eq.~(\ref{A7})
into $i/k_0$, while the terms within the first pair of brackets can be combined in view 
of the eigenvalue relation (\ref{e62}). These simplifications reduce Eq.~(\ref{A7}) to the form
\begin{eqnarray}
\label{A8}
&\int_0^R& r^2 h^{(1)}_1(k_0r^>)j_1(k_0r^<)j_1(\gn r) dr
={1\over (k_0^2-\gn^2)}\nonumber\\
&\times \Big\{&j_1(k_0r_0)Rh_1^{(1)}(k_0R)[Rj_1(\gn R)]^\prime
(1-k_0^2/\gn^2)\nonumber\\
&+&(i/k_0)j_1(\gn r_0)\Big\}.
\end{eqnarray}
When the integral (\ref{A8}) is substituted into the right hand side of Eq.~(\ref{A1}), the 
inner product attains its final form 
\begin{equation}
\label{A9}
(d(r,0),j_1(\gn r))=-4ik_0^2r_0^2\sqrt{4\pi\over 3}P_0{\gn^2\over (k_0^2-\gn^2)}
j_1(k_0r_0).
\end{equation}

The inner product $(j_1(\gn r),j_1(\gn r))$ involves, as definition (86)
indicates, the integral
$$\int_0^R dr j_1^2(\gn r) dr,$$
which can be evaluated exactly in closed form by means of the indefinite-integral 
identity \cite{Schiff}
$$\int x^2j_1^2(x)dx={x^3\over 2}[j_1^2(x)-j_0^2(x)j_2^2(x)].$$
This yields the following exact result for the inner product:
\begin{align}
\label{A10}
(j_1(\gn r),j_1(\gn r))={-2\over \gn}&\Big\{ xj_1(x)[xj_1(x)]^\prime\nonumber\\
&+ {x^3\over 2}[j_1^2(x)-j_0^2(x)j_2^2(x)]\Big\},
\end{align}
where $x=\gn R$. By substituting the explicit trigonometric forms for the spherical
Bessel functions of orders 0,1,2,
\begin{align*}
j_0(x)&={\sin x\over x}, \ \ j_1(x)={\sin x\over x^2}-{\cos x\over x},\nonumber\\
j_2(x)&=\left({3\over x^3}-{1\over x}\right)\sin x-{3\over x^2}\cos x,
\end{align*}
and performing simple algebraic manipulations,
we may reduce Eq.~(\ref{A10}) to the form
\begin{align}
\label{A11}
(j_1(\gn r),j_1(\gn r))={-1\over \gn}&\left[x
-\left(1-{4\over x^2}\right)\sin x\,\cos x\right.\nonumber\\
&-\left.{2\over x}\cos x-{2\over x^3}\sin x\right].
\end{align}
For large $|x|=|\gn|R$, only the first term in the square brackets in Eq.~(\ref{A11}) is
important,
\begin{equation}
\label{A12}
(j_1(\gn r),j_1(\gn r))\approx -R, \ \ \ \ |\gn|R>>1,
\end{equation}
which is the expression used in arriving at Eq.~(\ref{e104}).

\section{Evaluation of the First Sum in Eq.~(\ref{e112})}

Consider the contour integral
\begin{equation}
\label{B1}
I(\alpha) \equiv \oint_C {dz\over z(z-\alpha)\cos z},
\end{equation}
where the contour $C$ may be taken to be a circle of radius $N\pi$ in the complex-$z$ plane. 
Take $N$ to be a positive integer and $\alpha$ to be a finite complex number. 
In the limit $N\to\infty$, the integral
(\ref{B1}) must vanish, since the integrand goes to zero faster than $1/N^2$ while $dz$ grows only linearly
with $N$. But, by the residue theorem, the integral on the RHS is simply $2\pi i$ times
the sum of residues of the integrand at all of its poles in the finite complex plane. In the limit $N\to\infty$,
the integrand has only simple poles at 0, $\alpha$, and $(n-1/2)\pi$, $n=0,\pm 1,\pm 2, \ldots$, where
the residues are easily evaluated, and the following sum formula results:
\begin{widetext}
\begin{equation}
\label{B2}
0=2\pi i\left\{{-1\over \alpha}+{1\over \alpha\, \cos\alpha}-\sum_{n=-\infty}^\infty 
{1\over (n-1/2)\pi[(n-1/2)\pi -\alpha]\sin(n-1/2)\pi}\right\}.
\end{equation}
\end{widetext}
Note that $\sin(n-1/2)\pi = (-1)^{n-1}$. The infinite sum may be re-expressed as a one-sided sum
by relabeling $n$ by $(1-n^\prime)$ in the part of the sum that is over $n=-\infty$ to 0 and then
dropping the prime from $n^\prime$, as shown below: 
\begin{widetext}
\begin{eqnarray*}
\sum_{n=-\infty}^\infty {(-1)^n\over (n-1/2)[(n-1/2)\pi-\alpha]}
&=&\sum_{n=1}^\infty {(-1)^n\over (n-1/2)[(n-1/2)\pi-\alpha]}
+\sum_{n^\prime=1}^\infty{(-1)^{1-n^\prime}\over (1/2-n^\prime)[(1/2-n^\prime)\pi-\alpha]}\nonumber\\ 
&=&\sum_{n=1}^\infty {(-1)^n\over (n-1/2)}\left[{1\over (n-1/2)\pi-\alpha}-{1\over (n-1/2)\pi+\alpha}\right]\nonumber\\
&=&2\alpha \sum_{n=1}^\infty {(-1)^n\over (n-1/2) [(n-1/2)^2\pi^2-\alpha^2]}.\nonumber
\end{eqnarray*}
\end{widetext}
When this result is substituted into Eq.~(\ref{B2}), we obtain a closed-form expression for the sum 
needed in Eq.~(\ref{e112}), namely
\begin{equation}
\label{B3}
\sum_{n=1}^\infty {(-1)^n\over (n-1/2)\pi [(n-1/2)^2\pi^2-\alpha^2]} = {1\over 2\alpha^2}
\left(1-{1\over \cos\alpha}\right).
\end{equation}

\section{Cycle-Averaged Radiated Power}

When averaged over the fundamental oscillation period, $2\pi/\omega_0$,
the Poynting vector takes the form
\begin{equation}
\label{C1}
\vec S(\vr,t)=2c \ {\rm Re}\left[\vE(\vr,t)\times \vB^*(\vr,t)\right].
\end{equation}
The integral of the normal component of $\vec S$ over the surface of the sphere, $r=R$,
gives the total cycle-averaged power, $-dW/dt$, radiated by the sphere at time t,
\begin{equation}
\label{C2}
-{dW\over dt}=R\int d^2\Omega\, \vr\cdot \vec S|_{r=R}.
\end{equation}
The electric field $\vE$ is given by the expression (\ref{e105})), while the magnetic
field $\vB$, which obeys the Maxwell equation 
$$\vec D={i\over k_0}\grad\times \vB,$$
may be read off from Eq.~(\ref{e101}) for $\vec D$,
\begin{equation}
\label{C3}
\vB(\vr,t)={1\over 2ik_0}d(r,t)\vec LY_{10}.
\end{equation}
Using Eq.~(\ref{C1}) and a simple vector triple product rearrangement, we may write 
$\vec r\cdot \vec S$ as 
\begin{equation}
\label{C4}
\vec r\cdot \vec S=2c\, {\rm Re}(\vr\times\vE)\cdot\vB^*.
\end{equation}
In view of the form (\ref{e105}) for $\vE$, the vector identity 
$$\vr\times(\grad\times \vec A)=\grad(\vr\cdot \vec A)-\vec A-r{\partial\over\partial r}
\vec A,$$
and the operator identity $\vr\cdot\vec L=0$, it follows that 
\begin{equation}
\label{C5}
\vr\times \vec E=-{1\over 2k_0^2}{\partial\over\partial r}[re(r,t)] \vec LY_{10}(\vOm).
\end{equation}
With the help of Eqs.~(\ref{C5}) and (\ref{C3}) and noting that $\vec L^*=-\vec L$, we may reduce 
Eq.~(\ref{C4}) to the form
\begin{equation}
\label{C6}
\vec r\cdot \vec S= {\rm Re}\left[{ic\over 2k_0^3}d^*(r,t)[re(r,t)]^\prime
\vec LY_{10}(\vOm)\cdot\vec LY_{10}(\vOm)\right].
\end{equation}
Integrating Eq.~(\ref{C6}) over all solid angles with the help of a normalization integral for
spherical harmonics,
$$\int d^2\Omega\  \vec LY_{10}\cdot \vec LY_{10}=-2,$$
produces the following result for cycle-averaged power (\ref{C2}) at the spherical surface,
$r=R$:
\begin{equation}
\label{C7}
-{dW\over dt}=-2R\,{\rm Re}\left[{ic\over 2k_0^3}d^*(R,t)[R\ e(R,t)]^\prime\right].
\end{equation}

If we now use expansions (\ref{e102}) and (\ref{e106}) and make use of the first equality in Eq.~(\ref{e107}),
we may write Eq.~(\ref{C7}) as the double mode sum
\begin{align}
\label{C8}
-{dW\over dt}=-2R\,{\rm Re}\left\{ {ic\over 2k_0^3}\sum_m\sum_n\right. &d_m^* d_n j_1^*(\gm R)
{[Rj_1(\gn R)]^\prime\over (\gn^2/k_0^2)}\nonumber\\
&\left. \times e^{-\lambda^*t} e^{-\lambda_q t}\right\}.
\end{align}
Use of the eigenvalue equation (\ref{e62}) and the easily derived equality
$${{d\over d\beta}[\beta h^{(1)}_1(\beta)]\over h^{(1)}_1(\beta)}={\beta^2\over 
1-i\beta}-1,$$
helps us express the power $-dW/dt$ in the simple form
\begin{eqnarray}
\label{C9}
-{dW\over dt} &=& {c\beta^4\over k_0^4(1+\beta^2)}|\sum_n d_n j_1(\gn R)\exp(-\lnn t)|^2\nonumber\\
&=& {c\beta^2\over k_0^4}|d(r,t)|^2,
\end{eqnarray}
valid when $\beta>>1$.

\section{Evaluation of the Sum in Eq.~(\ref{e117})}

Sums of this form are most simply evaluated by integrating the associated function
\begin{equation}
\label{D1}
f(z)={\pi\over \sin\pi z}{\displaystyle{e^{i\alpha t/(\pi z-u)}\over (\pi z-u)}},
\end{equation}
where $u=\beta+(i/2)\ln \beta$, over a closed contour at infinity
that encloses all of the simple poles at all integers, $z=n$, $n=0,\pm1,\pm2,...$,
and the essential singularity at $z=u/\pi$ of the function $f(z)$. The integral over the 
contour vanishes, since the integrand $f(z)$ decays to 0 sufficiently rapidly as
$|z|\to \infty$. From the residue theorem of analytic function theory, then, the
sum of the residues at the poles and at the essential singularity must also vanish. Since
the residue at the pole $z=n$ is just 
$$(-1)^n {\displaystyle{e^{i\alpha t/(n\pi -u)}\over (n\pi -u)}},$$
the following sum formula results:
\begin{equation}
\label{D2}
\sum_{n=-\infty}^\infty(-1)^n {\displaystyle{e^{i\alpha t/(n\pi -u)}\over (n\pi -u)}}
=-\frac{\rm Residue}{z=u}\left[{\pi\over \sin\pi z} 
{\displaystyle{e^{i\alpha t/(z -u)}\over (z -u)}}\right].
\end{equation}

The residue at the essential singularity may be computed by expanding the exponential
in $f(z)$ in a power series and using the standard formula for the residue at a
pole of arbitrary order in each power-series term. This procedure leads to the 
following sum formula:
\begin{equation}
\label{D3}
\sum_{n=-\infty}^\infty(-1)^n {\displaystyle{e^{i\alpha t/(n\pi -u)}\over (n\pi -u)}}
=-\sum_{n=0}^\infty {(i\alpha t)^n\over (n!)^2} {d^n\over du^n}\left({1\over \sin u}
\right).
\end{equation}
The result (\ref{D3}), although exact, is not particularly useful since it trades one sum
for another. However, when $\beta>>1$, it can be expanded in a series of terms which 
decrease in magnitude as increasing positive powers of $1/\beta$. To do so, note that
we may write
$${1\over \sin u}={-2i \exp(iu)\over 1-\exp(2iu)}=-2i\sum_{m=0}^\infty\exp[i(2m+1)u],$$
and therefore 
\begin{equation}
\label{D5}
{d^n\over du^n}{1\over \sin u}=-2i\sum_{m=0}^\infty i^n (2m+1)^n\exp[i(2m+1)u].
\end{equation}
Substitution of the result (\ref{D5}) into Eq.~(\ref{D3}), followed by an interchange of the order
of the $m$ and $n$ sums on the right-hand side (RHS), leads to the following asymptotic sum formula:
\begin{align}
\label{D6}
\sum_{n=-\infty}^\infty(-1)^n {\displaystyle{e^{i\alpha t/(n\pi -u)}\over (n\pi -u)}}
=2i&\sum_{m=0}^\infty {\exp[i(2m+1)\beta]\over \beta^{m+1/2}}\nonumber\\
&\times\sum_{n=0}^\infty {[-(2m+1)\alpha t]^n\over (n!)^2} ,
\end{align}
where the value, $\exp(\pm i u)=\exp(\pm i\beta)(\beta)^{\mp1/2}$,
was used to replace $u$ in terms of $\beta$.
Since the $n$-sum on the RHS of Eq.~(\ref{D6}) is simply a power-series expansion of the Bessel function
$J_0(2\sqrt{(2m+1)\alpha t})$, Eq.~(\ref{D6}) reduces to a simpler form,
\begin{align}
\label{D7}
\sum_{n=-\infty}^\infty(-1)^n {\displaystyle{e^{i\alpha t/(n\pi -u)}\over (n\pi -u)}}
=2i&\sum_{m=0}^\infty {\exp[i(2m+1)\beta]\over \beta^{m+1/2}}\nonumber\\
&\times J_0(2\sqrt{(2m+1)\alpha t}).
\end{align}

\section{Approximate Evaluation of a Certain Sum }

When a function $g(n)$ changes slowly from one integer value of $n$ to the next, 
the sum $\sum_{n=-\infty}^0 (-1)^n g(n)$ may be evaluated by noting
that the sum of each successive pair of terms since they have opposite signs is 
approximately the same as the first derivative $g^\prime(n)$. We may thus write
\begin{equation}
\label{E1}
\sum_{n=-\infty}^0 (-1)^n g(n)\approx \sum_{m=-\infty}^0g^\prime(2m),
\end{equation}
where the sum is now only over even non-positive integers $n$: $n=2m$, $m=0,-1,-2,..$.
Since the sum on the right hand side of Eq.~(\ref{E1}) may be regarded, again approximately,
as an integral over $m$, we may write it as
\begin{equation}
\label{E2}
\sum_{n=-\infty}^0 (-1)^n g(n)\approx \int_{-\infty}^0 dm\, g^\prime(2m)={1\over 2}
\int_{-\infty}^0 dn\, g^\prime(n).
\end{equation}
Since the integrand is a total derivative with respect to the integration variable $n$,
its integral is trivial, and Eq.~(\ref{E2}) reduces to the simple form
\begin{equation}
\label{E3}
\sum_{n=-\infty}^0 (-1)^n g(n)\approx {1\over 2}g(0),
\end{equation}
since $g(-\infty)=0$ as a necessary consequence of the convergence of the original sum.

Use of this general result 
immediately proves the following sum formula:
\begin{equation}
\label{E4}
\sum_{n=-\infty}^0(-1)^n {\displaystyle{e^{i\alpha t/(n\pi -u)}\over (n\pi -u)}}
\approx -{1\over 2} {\displaystyle{e^{-i\alpha t/u}\over u}}.
\end{equation}
Noting that $u=\beta+(i/2)\ln\beta$ and $1/u\approx 1/\beta-(i/2)(\ln\beta)/\beta^2$,
valid for large $\beta$, the preceding result is approximately the same as
\begin{equation}
\label{E5}
\sum_{n=-\infty}^0(-1)^n {\displaystyle{e^{i\alpha t/(n\pi -u)}\over (n\pi -u)}}
\approx -{1\over 2} {\displaystyle{e^{-i\alpha t/\beta}e^{-\alpha t\ln\beta/(2\beta^2)}
\over \beta}}.
\end{equation}
Subtracting this result from the first term of Eq.~(\ref{D7}), which for large $\beta$
gives the sum over {\it all} integer values of $p$ for the same summand, taking the 
squared modulus of the resulting difference, and then keeping only the two most 
significant powers of $1/\sqrt{\beta}$ yield the result (\ref{e119}) for the 
corresponding sum over only positive integral values of $n$.


\begin{thebibliography}{99}
\bibitem{footnote1} Differences by a factor 2 in the definitions of the complex 
polarization amplitudes and by a factor $4\pi$ because of our use of
rationalized, rather than ordinary, Gaussian units 
must be accounted for in order to use the radiation rate given in 
Ref.~\cite{Jackson1}.
\end{thebibliography}

\end{document}